\documentclass[prd,tightenlines,nofootinbib,superscriptaddress]{revtex4}

\usepackage{amsfonts,amssymb,amsthm,bbm}

\usepackage{amsmath}
\DeclareMathOperator\arctanh{arctanh}
\DeclareMathOperator\arcsinh{arcsinh}

\usepackage{hyperref}

\usepackage{color,psfrag}
\usepackage[dvips]{graphicx}

\usepackage{tikz}
\usetikzlibrary{calc}
\usetikzlibrary{decorations.pathmorphing}
\usetikzlibrary{shapes.geometric}
\usetikzlibrary{arrows,decorations.markings}
\usetikzlibrary{patterns}

\newcommand{\C}{{\mathbb C}}
\newcommand{\N}{{\mathbb N}}
\newcommand{\R}{{\mathbb R}}

\newcommand{\cA}{{\mathcal A}}

\newcommand{\cG}{{\mathcal G}}

\newcommand{\cH}{{\mathcal H}}
\newcommand{\cM}{{\mathcal M}}
\newcommand{\cN}{{\mathcal N}}

\newcommand{\cT}{{\mathcal T}}
\newcommand{\cV}{{\mathcal V}}

\newcommand{\cC}{{\mathcal C}}
\newcommand{\cS}{{\mathcal S}}

\newcommand{\cI}{{\mathcal I}}

\newcommand{\vecc}{\overrightarrow}

\newcommand{\SU}{\mathrm{SU}}

\newcommand{\SO}{\mathrm{SO}}
\renewcommand{\O}{\mathrm{O}}
\newcommand{\U}{\mathrm{U}}

\newcommand{\be}{\begin{equation}}
\newcommand{\ee}{\end{equation}}
\newcommand{\beq}{\begin{eqnarray}}
\newcommand{\eeq}{\end{eqnarray}}
\newcommand{\bes}{\begin{eqnarray}}
\newcommand{\ees}{\end{eqnarray}}

\newcommand{\mat} [2] {\left ( \begin{array}{#1}#2\end{array} \right ) }

\newcommand{\su}{{\mathfrak su}}

\newcommand{\la}{\langle}
\newcommand{\ra}{\rangle}

\newcommand{\tr}{{\mathrm{Tr}}}
\newcommand{\f}{\frac}

\def\nn{\nonumber}
\def\pp{\partial}
\newcommand{\w}{\wedge}

\def\vphi{\varphi}
\def\eps{\epsilon}

\newcommand{\id}{\mathbb{I}}

\def\vx{\vec{x}}
\def\vv{\vec{v}}
\def\vJ{\vec{J}}

\def\vN{\vec{N}}
\def\vsigma{\vec{\sigma}}
\def\arr{\rightarrow}

\def\hn{\hat{n}}

\def\tT{\tilde{T}}

\def\tcG{\widetilde{\cG}}

\def\tt{\tilde{t}}
\def\dd{\mathrm{d}}
\def\vX{\vec{X}}
\def\vcC{\vec{\cC}}
\def\hT{\hat{T}}
\def\hcT{\hat{{\cT}}}

\begin{document}

\title{Probing the Shape of Quantum Surfaces: the Quadrupole Moment Operator}

\author{{\bf Christophe Goeller}}\email{christophe.goeller@ens-lyon.fr}
\affiliation{Univ Lyon, Ens de Lyon, Universit\'e Claude Bernard Lyon 1, CNRS, 
Laboratoire de Physique, F-69342 Lyon, France}
\affiliation{Perimeter Institute for Theoretical Physics, 31 Caroline St N, Waterloo ON, Canada N2L 2Y5}

\author{{\bf Etera R. Livine}}\email{etera.livine@ens-lyon.fr}
\affiliation{Univ Lyon, Ens de Lyon, Universit\'e Claude Bernard Lyon 1, CNRS, 
Laboratoire de Physique, F-69342 Lyon, France}

\date{\today}

\begin{abstract}


The standard toolkit of operators to probe quanta of geometry in loop quantum gravity consists in area and volume operators as well as holonomy operators. New operators have been defined, in the $\U(N)$ framework for intertwiners, which allow to explore
the finer structure of quanta of geometry.
However these operators do not carry information on the global shape of the intertwiners.
Here we introduce dual multipole moments for continuous and discrete surfaces, defined through the normal vector to the surface, taking special care to maintain parametrization invariance.
These are raised to multipole operators probing the shape of quantum surfaces.
Further focusing on the quadrupole moment, we show that it appears as the Hessian matrix of the large spin Gaussian approximation of coherent intertwiners, which is the standard method for extracting the semi-classical regime of spinfoam transition amplitudes.
This offers an improvement on the usual loop quantum gravity techniques, which mostly focus on the volume operator, in the perspective of modeling (quantum) gravitational waves as shape fluctuations waves propagating on spin network states.

\end{abstract}

\maketitle
\tableofcontents

\section*{Introduction}

Loop quantum gravity defines a framework for a canonical approach to the quantization of general relativity and describes quantum states of geometry evolving in time (see \cite{Bodendorfer:2016uat} for a recent review or \cite{Thiemann:2007zz} for a complete textbook). Despite tremendous progress over the past decade on both theoretical and phenomenological fronts, the fundamental theory still faces tough issues of the precise definition of an anomaly-free dynamics, the fate of diffeomorphism invariance at the quantum level and the related definition of a quantum observer in a quantum geometry, and the explicit implementation of a coarse-graining flow from Planck scale quantum geometries to larger scales in order to extract rigorous renormalized quantum gravity correction to gravitational physics. All of these require the development of mathematical tools to probe, handle and analyze the quantum states of geometry defined by loop quantum gravity as (superpositions of) spin network states. In this spirit, we would like to extend the loop quantum gravity toolkit by introducing new operators acting on spin networks, which probe the shape of elementary quanta of volume and more generally of quantum surfaces. The multipole moment operators, which we define in this paper, allow to analyze the global shape of quanta of geometry beyond the usual area and volume and should be relevant to identifying and modeling gravitational waves in loop quantum gravity as shape fluctuations waves propagating on spin network states.

More precisely, those canonical states of geometry are endowed with an intrinsic discrete structure and implement a discrete spectra for geometric observables such as areas and volumes \cite{Rovelli:1994ge,Ashtekar:1997fb,Rovelli:1998gg}. As mathematical objects, they are defined on graphs, as networks of vertices linked by edges all dressed with algebraic objects from the theory of representations of the Lie group $\SU(2)$. To edges are attached spins, that is $\SU(2)$-representations, which define the quanta of area carried by the edge. And to vertices are attached intertwiners, that is $\SU(2)$-invariant tensor states, which define the quanta of volume located at the vertex. This provides spin networks with a natural interpretation as discrete geometries, best understood as twisted geometries \cite{Freidel:2010aq,Dupuis:2012yw,Freidel:2013bfa}. Thus, on a given spin network, a vertex defines an elementary block of volume. Each edge attached to this vertex defines a surface between that elementary block of space and its neighboring vertex located at the other end of the edge. Together all these edges around a vertex define the dual surface surrounding the elementary volume. More generally, putting several vertices together (considering a connected subgraph) defines a bounded region of the spin network. This region's boundary is a quantum surface defined by the edges linking that region to the exterior \cite{Feller:2017ejs}. These (quantum) surfaces are defined through the normal vector to the surface, which encodes both  intrinsic and extrinsic geometries of the surface embedded in the 3d space.  Here we will define multipole moment observables in terms of the normal vectors and raise them to multipole operators acting on intertwiners (and more generally on quantum surfaces). 

These multipole operators provide a further tool to probe the geometry of spin networks, on top of the standard area operators, volume operators playing a crucial role in the quantization of the Hamiltonian constraints {\`a la Thiemann} \cite{Thiemann:1996aw,Thiemann:1996av,Ashtekar:1997fb,Brunnemann:2004xi,Giesel:2005bk,Giesel:2005bm,Alesci:2014aza} and $\U(N)$-operators which define basic exchange of quanta of areas between spin network edges \cite{Girelli:2005ii,Freidel:2009ck,Freidel:2010tt,Livine:2013tsa,Girelli:2017dbk}. We have in mind applications to the analysis of the loop quantum gravity dynamics, especially the propagation of gravitational waves seen as deformations of the geometry, and possibly to the coarse-graining of spin networks which aims to understanding the emergence of classical geometry and curvature from the quantum structures of loop quantum gravity.

\medskip

In this paper, we will start by introducing the dual multipole moments for a classical 2d surface, as integrals of tensor products of the normal vector field. We will compare them to the definition of the usual multipole moments and will insist on the reparametrization invariance of these surface integrals. Focusing on the (dual) quadrupole moment, it allows to distinguish ellipsoid shapes from the straightforward spherical shape. We will introduce the quadrupole moment observable for discrete surfaces, describe the transformations that live it invariance and illustrate these definitions in section 2 with the example of tetrahedra. In section 3, we will show that the quadrupole moment matrix arises as the Hessian matrix controlling the shape of coherent intertwiners in their large spin asymptotics. These coherent intertwiners \cite{Livine:2007vk,Freidel:2009nu,Freidel:2009ck,Freidel:2010tt,Livine:2013tsa,Girelli:2017dbk,Bonzom:2012bn} have become essential to the construction of spinfoam models for the dynamics of loop quantum gravity and the study of their semi-classical behavior \cite{Freidel:2007py,Livine:2007ya,Barrett:2009mw,Barrett:2010ex}. As a mathematical tool, they are also applied to the construction of a holomorphic representation of spin network states \cite{Freidel:2009nu,Borja:2010rc,Livine:2011gp,Bonzom:2011nv,Alesci:2016dqx}. Finally, we will conclude this paper with the definition of the quadrupole moment operator acting on intertwiners. In particular, we will underline an unexpected factor, inverse of the dimension $(2j+1)$ of a spin representation, which implies that the operator is not polynomial in the $\su(2)$ generators but nevertheless ensures the reparametrization invariance of the observable in the continuum limit.

\section{Dual Multipole Moment for a Surface}

Intertwiner states, living at the vertices of spin networks, have the semi-classical interpretation of convex polyhedra embedded in the flat 3d Euclidean space \cite{Freidel:2009ck,Livine:2013tsa,Bianchi:2010gc}. More precisely, considering a vertex to which $N$ links are attached, the semi-classical data encoded by an intertwiner state is a 3d vector $\vN_{i}\in\R^{3}$ for each link $i=1..N$. The $\SU(2)$-invariance of the intertwiner state translates into a closure constraint, $\sum_{i}\vN_{i}=0$, which ensures (by a theorem of Minkowski) the existence of a unique convex polyhedron with $N$ faces, each face dual to a link with the normal vector to a face given precisely by $\vN_{i}$. In particular, the norm $|\vN_{i}|$ gives the area of the corresponding face. An explicit reconstruction algorithm is explicited in \cite{Bianchi:2010gc}. At the quantum level, the area, i.e. the normal vector norm, is given by the spin carried by the corresponding spin network link.

Beside the area of the polyhedron's faces (and the total boundary area around the vertex), the usual observables used to probe intertwiner states are the volume \cite{Bianchi:2010gc,Barbieri:1997ks} and the scalar products between normal vectors encoding the dihedral angles between the polyhedron faces (and their refinement in the so-called $\U(N)$ observables for intertwiners) \cite{Freidel:2010tt,Livine:2013tsa,Dupuis:2010iq}. What is still missing are multipole moment observables, which probe the global shape of the intertwiner and corresponding semi-classical polyhedron. Here we remedy this gap by studying the dual multipole moments for a 2d surface embedded in the flat 3d space and applying them to polyhedra and quantum intertwiner states.

\subsection{Multipole moments vs. ``Dual'' Multipole Moments}
\label{multipole}

Let $\cS$ be a two-dimensional surface  embedded in the 3d flat Euclidean space $\R^{3}$.
%
%
We parametrize the surface with two coordinates $(u,v)$ defining the surface points $\vx(u,v)\in\R^{3}$. We will use the indices $A,B=0,1$ for the surface coordinates with $u_{0}=u$ and $u_{1}=v$, while indices $a,b,c$ label 3d vector components. 
The metric induced on the surface is:
\be
\dd s^{2}
\,=\,
g_{AB}\dd u^{A}\dd u^{B}
\,=\,
\left(\pp_{u}\vx\cdot\pp_{u}\vx\right)\,\dd u^{2}
+2\,\left(\pp_{u}\vx\cdot\pp_{v}\vx\right)\,\dd u\dd v
+\left(\pp_{v}\vx\cdot\pp_{v}\vx\right)\,\dd v^{2}
\,,
\qquad
g_{AB}=\pp_{A}\vx\cdot\pp_{B}\vx\,.
\ee
The infinitesimal area element is:
\be
\dd A
\,=\,
\sqrt{\det g}\,\dd u\dd v
\,,
\qquad
\sqrt{\det g}
\,=\,
\sqrt{\left(\pp_{u}\vx\cdot\pp_{u}\vx\right)\left(\pp_{v}\vx\cdot\pp_{v}\vx\right)
-\left(\pp_{u}\vx\cdot\pp_{v}\vx\right)^{2}}
\,=\,
|\pp_{u}\vx \w \pp_{v}\vx|
\,.
\ee
The moments of the surface $\cS$ are surface integrals of polynomials in the coordinates. The $p$-th moment $\cI_{p}$ is a tensor of rank $p$ (with $3^{p}$ components),
\be
\cI_{p}^{a_{1}..a_{p}}=
\int_{\cS}\dd u \dd v\, \sqrt{\det g}\,\prod_{i=1}^{p}x^{a_{i}}
\,.
\ee
At the zeroth order, the monopole gives the total surface area:
\be
\cI_{0}
\,=\,
\int_{\cS} \dd u\dd v\,\sqrt{\det g}
\,=\,
\int_{\cS} \dd A
\,=\,
\cA\,.
\ee
The dipole moment takes the mean of the positions and defines the center of mass of the surface:
\be
\cI_{1}^{a}
\,=\,
\int_{\cS} \dd u\dd v\,\sqrt{\det g}\,x^{a}
\,=\,
\vX\,.
\ee
The quadrupole moment is defined as a 3$\times$3 symmetric matrix\footnotemark{} and probes the basic shape of the surface:
\be
\cI_{2}^{ab}
\,=\,
\int_{\cS} \dd u\dd v\,\sqrt{\det g} \,x^{a}x^{b}
\,.
\ee
\footnotetext{
We  usually  define the traceless part of the quadrupole matrix, $\tilde{t}^{ab}=\cI_{2}^{ab}-\f13\tr \cI_{2} \,\id_{3}$, which vanishes for a 2-sphere.
}
For a closed surface centered on the origin, i.e. such that $\cI_{1}^{a}=0$, the quadrupole moment measures the deviation of the surface from a perfect sphere. Indeed, the sphere has a vanishing quadrupole moment. Then the three eigenvalues of a non-vanishing quadrupole moment give the three radii of the best  ellipsoid approximation of the surface (see in appendix \ref{app:ellipsoid} for more details).
Higher order polynomials and moments will explore the finer structure of the surface.

\bigskip

We have reviewed above the standard moments for a surface. However, in the context of loop quantum gravity, the geometry of a surface is not defined through the positions $\vx$ but through the normal to the surface.
In differential terms, the surface normal is defined as:
\be
\vN=\pp_{u}\vx \w \pp_{v}\vx,
\qquad
N= |\vN|,
\qquad
\hn=\f{\vN}{N}
\,.
\ee
Its norm gives the infinitesimal area element:
\be
N= \sqrt{\det g}
\,,\quad
dA
\,=\,
N\,\dd u\dd v\,.
\ee
Assuming that we are provided with the distribution of normal vectors instead of the position vectors, it is natural to define the moment of the normal vectors. This leads us to introduce  dual moments for the surface. The essential point here is that the unit normal $\hn$ is invariant under reparametrization (it is a scalar), so we define the dual moments as the integral means of the powers of the unit normal,
\be
\cM_{r}^{a_{1}..a_{r}}
\,=\,
\int_{\cS}\dd u \dd v\,N \,\prod_{i=1}^{r}\hat{n}^{a_{i}}
\,=\,
\int_{\cS}\dd u \dd v\,N^{1-r} \,\prod_{i=1}^{r}{N}^{a_{i}}
\,.
\ee
All these integrals are properly weighted so they are invariant under surface reparametrization and do not depend on the specific coordinate system chosen to define them.
The dual monopole moment is simply the area once again:
\be
\cM_{0}
\,=\,
\int_{\cS} \dd u \dd v\,N
\,=\,
\cI_{0}=\cA
\ee
The dual dipole is the closure vector (or defect) usually used in loop quantum gravity and twisted geometry:
\be
\cM_{1}^{a}
\,=\,
\int_{\cS} \dd u \dd v\,N \,n^{a}
\,=\,
\int_{\cS} \dd u \dd v\,N^{a}
\,.
\ee
When the surface is closed, the closure vector vanishes,
$$
\cM_{1}^{a}=\int_{\cS} \dd\big{(}\epsilon^{abc}x^{b}\,\dd x^{c}\big{)}
\quad\underset{\cS\textrm{ closed}}{=}\quad 0\,.
$$
The dual quadrupole is the main object of interest in the present work:
\be
T^{ab}
\,\equiv\,
\cM_{2}^{ab}
\,=\,
\int_{\cS}\dd u \dd v\,N \,n^{a}n^{b}
\,=\,
\int_{\cS} \dd u \dd v\,\f1N \,N^{a}N^{b}\,.
\ee
The trace of this dual quadrupole tensor is simply the surface area\footnotemark:
\be
\tr\,T=\int _{\cS} \dd u \dd v\,N=\cA\,.
\ee
\footnotetext{
We can introduce the traceless part of the dual quadrupole moment, removing the information already contained by the lower order moments:
$$
\tT^{ab}=T^{ab}-\f13\tr T \,\id_{3}\,.
$$}
In the present work, we will focus on the dual quadrupole moment, which defines an ellipsoid approximation of the surface, and on its quantization and action on coherent intertwiners and quantum surfaces.
Since the surface normals rotate as vectors under 3d rotations, one can rotate the surface in order to diagonalize this dual quadrupole tensor:
\be
T={}^{t}O \mat{ccc}{\mu_{1} & & \\ &\mu_{2}& \\ &&\mu_{3} }O\,,\quad
O\in\SO(3)
\ee
%
The three eigenvalues $\mu_{a}$ encode the  shape of the surface. For a closed surface with spherical topology, its dual quadrupole tensor defines a unique ellipsoid approximation\footnotemark{}, such that this ellipsoid yields the exact same dual quadrupole moment. The rotation $O$ defines the principal axes of the ellipsoid, and we can extract the three ellipsoid radii from these three quadrupole matrix eigenvalues. This new ellipsoid approximation defined by the dual quadrupole has exactly the same area as the original surface. More details are in given in appendix \ref{app:dualellipsoid}.
\footnotetext{
Generically, this ellipsoid approximation is a priori different from the one obtained from the usual quadrupole moment $\cI_{2}$.}
%

\subsection{Quadrupole Moment for Discrete Surfaces}

In the context of loop quantum gravity, surfaces are intrinsically discrete and be thought, in the semi-classical regime, as triangulated surface. Let us consider a piecewise-linear closed surface, which we think of as a polyhedron with $P$ faces. We describe this discrete surface by the normal vector $\vN_{i}\in\R^{3}$ to each face $i=1..P$. The area of each face is given by the norm $N_{i}=|\vN_{i}|$, so that the total surface area is simply the sum of the norms of the normal vectors:
\be
\cA=\cM_{0}=\sum_{i}^{P}N_{i}
\,.
\ee
The dipole moment is the sum of the normal vectors,
\be
\cC^{a}=\cM_{1}^{a}=\sum_{i}^{P}N_{i}^{a}\,.
\ee
For a closed surface, this vector vanishes, that's the closure constraint, $\cC^{a}=0$. Reciprocally, starting with $P$ vectors satisfying the closure constraint, a theorem by Minkowski ensures the existence of a unique convex polyhedron such that the normal vectors to its faces are exactly the initial vectors. This fact is at the heart of the interpretation of spin network states in loop quantum gravity as discrete twisted geometries \cite{Bianchi:2010gc}.

We define the (dual) quadrupole moment as a straightforward discretization of the integral for differentiable surfaces:
\be
T^{ab}=\sum_{i=1}^{P}\f1{N_{i}}N_{i}^{a}N_{i}^{b}
\,,\qquad
\tr\,T=\cA\,.
\label{eq:discrete_dual_moment_def}
\ee
The key remark at this point is that this expression is not a polynomial in the face normals. Indeed, from the perspective of studying polynomial observables of the face normals, it would have been more natural to consider $\sum_{i=1} N_{i}^{a}N_{i}^{b}$. However, the inverse norm factor $1/N_{i}$ is needed to match the correct definition in the continuum, which is invariant under re-parametrization. Actually, it turns out that this factor is essential to ensure a correct continuum limit under surface refinement as the number of faces is sent to infinity $P\arr\infty$, which highlights the key interplay between continuum limit and diffeomorphism invariance in quantum gravity \cite{inprep}.

We can similarly define higher order moments of a discrete (closed) surface, with the appropriate inverse norm factors:
\be
\cM_{r}^{a_{1}..a_{r}}=\sum_{i=1}^{P}\f1{N_{i}^{r-1}}N_{i}^{a_{1}}..N_{i}^{a_{r}}\,.
\ee
These inverse norm factors will clearly be an issue for the quantization of these observables into operators acting on intertwiners (and quantum surface states). Preserving the classical Lie algebra commutators will very likely fix the operator ordering and quantization ambiguities. We will discuss this issue for the definition of the quadrupole moment operator later in section \ref{operator}, but we postpone its general analysis  to future work.

\subsection{Scalar products, Gram matrix and $\O(N)$ deformations}

Let us look more closely at the the quadrupole moment $T$. As a 3$\times$3 matrix, its three eigenvalues are entirely determined by the traces of its powers, $\tr T$, $\tr T^{2}$, $\tr T^{3}$:
\be
\tr\,T=\sum_{i}N_{i}=\cA
\,,\quad
\tr\,T^{2}=\sum_{i,j}\f1{N_{i}N_{j}}\big{(}\vN_{i}\cdot\vN_{j}\big{)}^{2}
\,,
\quad
\tr\,T^{3}=\sum_{i,j,k}\f1{N_{i}N_{j}N_{k}}\big{(}\vN_{i}\cdot\vN_{j}\big{)}\big{(}\vN_{j}\cdot\vN_{k}\big{)}\big{(}\vN_{k}\cdot\vN_{i}\big{)}\,.
\label{eq:TrT_TrT2_TrT3_expression}
\ee
These traces are invariant under 3d rotations. More precisely, they are combinations, sums and products, of the scalar product observables $\vN_{i}\cdot\vN_{j}$, which give the dihedral angles between the faces. While the scalar products between neighboring faces probe the local shape of the shape, the traces $\tr\,T^k$ describe the global shape of the surface.

In light of these relations, it seems natural to introduce the  $P$$\times$$P$ Gram matrix
\be
\cG_{ij}=\vN_{i}\cdot\vN_{j}
\,.
\ee
The closure constraint $\sum_{i}\vN_{i}=0$ provides a null vector to the Gram matrix, $\sum_{j} \cG_{ij}v_{j}=0$ with $v_{j}=1$ for all $j$'s.
To compare with the quadrupole moment matrix, it is better to introduce a modified normalization for the Gram matrix:
\be
\tcG_{ij}=\f1{\sqrt{N_{i}}}\vN_{i}\cdot\f1{\sqrt{N_{j}}}\vN_{j}
\,,
\qquad
\cG_{ii}=N_{i}
\,.
\ee
This awkward normalization of the normal vector allows to match the inverse norm factors for the power traces:
\be
\tr \,\tcG=\sum_{i}N_{i}=\cA=\tr\,T
\,,\qquad
\tr \,\tcG^{k}=\tr \,T^{k}\quad\forall k\ge 1\,.
\ee
%
In particular, this implies that $T$ and $\tcG$ have the same eigenvalue spectrum (excluding the null vectors of the Gram matrix). In this sense, the dual  quadrupole moment $T$ contains the same information as the renormalized Gram matrix $\tcG$. But on the one hand, one is looking at the $\O(3)$-invariant data contained in $T$ and, on the other hand, the $\O(P)$-invariant data in $\tcG$. The $\O(3)$ transformations are the 3d rotations, but what is the role of those $\O(P)$ transformations?

A $\O(P)$ transformation mixes the normal vectors of all the faces,
\be
\f{N_{i}^{a}}{\sqrt{N_{i}}}
\,\longmapsto\,
\f{\cN_{i}^{a}}{\sqrt{\cN_{i}}}=
\sum_{j}^{P}\Omega_{ij}\f{N_{j}^{a}}{\sqrt{N_{j}}}
\,,
\qquad
{}^{t}\Omega\Omega=\id_{P}
\,,\quad
\Omega\in\O(P)\,.
\ee
It is a non-linear transformation of the normal vectors. In particular, we get for their norm:
\be
N_{i}
\,\longmapsto\,
\cN_{i}=\sum_{j,k}\Omega_{ij}\Omega_{ik}\f{\vN_{j}\cdot\vN_{k}}{\sqrt{N_{j}N_{k}}}
\,,
\ee
which depends on all the initial normal vectors.
Since the Gram matrix simply transforms by conjugation, $\cG\,\mapsto\,\Omega \,\tcG \,\Omega^{-1}$, these $\O(P)$ transformations leave its spectrum invariant, as well as the spectrum of the quadrupole moment $T$. These $\O(P)$ transformations actually allow to explore all the polyhedra with the same quadrupole tensor (up to 3d rotations). Thus, this $\O(P)$ group structure is specially interesting if one wants to modify the discrete surface without changing its shape (at least, at the level of the quadrupole moment) or if one wants to average over the ensemble of all discrete surfaces with the same shape (i.e. quadrupole moment up to 3d rotations).

The role of these $\O(P)$ transformations is very similar to the $\U(P)$ transformations acting on (framed) polyhedra introduced in \cite{Freidel:2010tt,Livine:2013tsa}, which allow to explore all the space of (convex) polyhedra with fixed total area. There are two differences. First, the unitary transformations act on framed polyhedra, which have an extra data of a phase attached to each face. Second, the present orthogonal transformations leave the whole quadrupole moment invariant (up to 3d rotations) and  not only the total area (which is the trace of the quadrupole moment matrix).

It is tempting to interpret them as the discrete equivalent of surface re-parameterizations, since they both leave the quadrupole moment invariant. However, in order to consolidate this interpretation, we should extend them to transformations that leave all the discrete multipole moments invariant. We postpone a deeper analysis of this structure to a future investigation.

\subsection{Quadrupole and Volume Observable(s)}

The quadrupole moment allows to define a volume observable for the surface, as the volume of the ellipsoid approximation to the surface., which we can expressed in terms of the ellipsoid radii, computed from the quadrupole moment eigenvalues. The ellipsoid approximation has the same area of the original surface but the actual bounded volume is a priori different.
This is different from the usual volume observables and operators used in loop quantum gravity. Indeed, the typical volume observable is constructed from the mixed product of triplets of normal vectors:
\be
U\propto\sum_{i,j,k}\eps^{ijk}\vN_{i}\cdot(\vN_{j}\wedge\vN_{k})
\,,
\ee
where $\eps^{ijk}$ defines an orientation over triplets of faces. In the case of $P=4$ faces, that is for tetrahedra, this gives exactly the square of the tetrahedron volume (up to a orientation sign and the appropriate numerical factor). Volume operators in loop quantum gravity usually\footnotemark{} start with the observable $U$ and slightly differ in their definition of the orientation of triplets, ordering ambiguities and how to take the square-root to get a volume back.
\footnotetext{
There is nevertheless another very compelling proposal. Indeed, for a higher number of faces $P \ge 5$, the observable $U$ deviates from the actual squared volume of the polyhedra. Instead, one could simply take the volume of the polyhedra, as suggested in \cite{Bianchi:2010gc}. However, the reconstruction of the (convex) polyhedron from the normal vectors is not a simple algorithm but is realized through an optimization method. This means that there is no classical formula that can be straightforwardly quantized. So, defining a volume operator at the quantum level can only work if dealing exclusively with classical polyhedron in a coherent state quantization scheme \cite{Bianchi:2010gc}.
}

The observable $U$ does not seem related to the ellipsoid volume defined by the (dual) quadrupole moment. There is however a more subtle link. We can compute the determinant of the quadrupole moment matrix\footnotemark{}:
\be
\det T
=
\f16\sum_{i,j,k}\f1{N_{i}N_{j}N_{k}}\big{|}
\vN_{i}\cdot(\vN_{j}\wedge\vN_{k})
\big{|}^2
=
\f16\sum_{i,j,k}{N_{i}N_{j}N_{k}}\big{|}
\hn_{i}\cdot(\hn_{j}\wedge\hn_{k})
\big{|}^2
\ee
\footnotetext{Since $T$ is a 3$\times$3 matrix, its determinant $\det T$ is related algebraically to its power traces:
\be
6\det\,T
\,=\,
(\tr \,T)^3-3(\tr\,T)(\tr\,T^2)+2\tr\,T^3
\,.
\nn
\label{eq:detT_expression}
\ee
}
Since the matrix elements $T^{ab}$ have the dimension of areas, this determinant $\det \, T$ has the dimension of a squared volume, i.e. the same physical dimension as $U$. So one can consider $\det\,T$ as defining a legitimate alternative to $U$ for loop quantum gravity.
Moreover, when working with tetrahedra, i.e. in the $P=4$ case, $\det\,T$ gives exactly the fourth power of the tetrahedron volume $U^{2}\propto V^{4}$, up to an area factor. This will be investigated in more details in the next section dedicated to the $P=4$ case.

At the end of the day, it would be interesting to explore further if $\det\,T$ provides a relevant volume observable for the kinematics and dynamics of spin networks in loop quantum gravity.

\section{The Shape of a Tetrahedron}

In this section, we apply the above construction to the simplest polyhedra we know and compute the (dual) quadrupole moment of  tetrahedra. Thus considering the $P=4$ case with four normal vectors, we focus on tetrahedra with fixed triangle areas, i.e. keeping the norms of the four normal vectors $N_{i}$ fixed, and study how the quadrupole moment probes the shape of the tetrahedra. We give explicit formulas and numerics in the iso-area case when the four normals have the same norm.

\subsection{The dual quadrupole moment of the tetrahedron}

A tetrahedron in $\R^{3}$ is a 4-simplex, consisting in four points linking by six edges organized in four triangles. It is fully determined by the positions of the four points, that is 12 parameters. If we factor out translations in $\R^{3}$ and 3d rotations, the tetrahedron's shape is determined by 6 parameters. The usual parametrization is to consider the lengths of the six edges of the tetrahedron, satisfying the relevant triangle inequalities (and positivity of the Cayley determinant).

In the context of loop quantum gravity and spinfoam models, the geometry is formulated in terms of an ``area Regge calculus'' and the tetrahedron is defined through the normal vectors of its four triangles, $\vN_{i}$ for $i$ running from 1 to 4. These normals satisfy the closure constraint, $\sum_{i}\vN_{i}=0$. It is possible to uniquely reconstruct the tetrahedron (up to translations in 3d) from these normal vectors\footnotemark. This is the simplest case of the Minkowski reconstruction theorem for convex polyhedra.
\footnotetext{
The edge vector $e_{ij}$ between the two triangles $i$ and $j$ is proportional to the vector product between the normal vectors to those triangles, $e_{ij}\propto\vN_{i}\w \vN_{j}$. The proportionality factor is the square-root of the squared volume observable $U=|\vN_{1}\cdot(\vN_{2}\w \vN_{3})|$ (up to a $\f29$ numerical factor). We are left with a sign ambiguity, fixed by the relative orientation of each edge with the triangles to which it belongs.
}

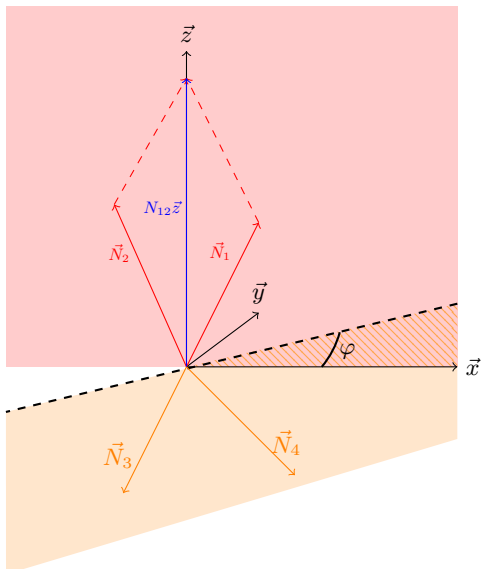
\begin{figure}
	\begin{center}
		\begin{tikzpicture}[scale=1.2]
		
		\coordinate (O) at (10,2);
		\coordinate (x) at (13,2);
		\coordinate (z) at (10,5.5);
		\coordinate (y) at (10.8,2.6);
		
		\coordinate (N1) at (10.8,3.6);
		\coordinate (N2) at (9.2,3.8);
		\coordinate(N2') at (10,5.2);
		
		\coordinate (N3) at (9.3,0.6);
		\coordinate (N4) at (11.2,0.8);
		
		\coordinate (varphiR) at (13,2.7);
		\coordinate (varphiL) at (8,1.5);
		
		\fill[color=orange!20] (varphiL)--(varphiR) --+(0,-1.5)--+(-5,-3)--cycle;
		\fill[color=red!20] (x)--+(0,4)--+(-5,4)--+(-5,0)--cycle;
		\fill[pattern=north west lines, pattern color=orange!70] (x)--(O)--(varphiR)--cycle;
		
		
		\draw[dashed, thick] (varphiR)--(varphiL);
		\draw[thick] (11.5,2) arc (-40:-15:1); \draw (11.6,2) node[above right]{$\varphi$};

		
		\draw[->] (O)-- node[pos=1,right]{$\vec{x}$}(x);
		\draw[->] (O)-- node[pos=1,above]{$\vec{y}$}(y);
		\draw[->] (O)-- node[pos=1,above]{$\vec{z}$}(z);
		
		\draw[red,->] (O)-- node[scale=0.7,pos=0.7,above left]{$\vec{N}_1$}(N1);
		\draw[color=red,->] (O)-- node[scale=0.7,pos=0.7,left]{$\vec{N}_2$}(N2);
		\draw[dashed,color=red,->] (N1)--  (N2');
		\draw[dashed,color=red,->] (N2)--  (N2');
		
		\draw[blue,->] (O)-- node[pos=0.55,left,scale=0.7]{$N_{12} \vec{z}$}(N2');
		
		\draw[orange,->] (O)-- node[pos=0.7,left]{$\vec{N}_3$}(N3);
		\draw[orange,->] (O)-- node[pos=0.7,right]{$\vec{N}_4$}(N4);
			
		
		\end{tikzpicture}
	\end{center}
	\caption{Dual representation of the tetrahedron in terms of the normal vectors to its faces. 3d rotations are fixed by assuming that $\vec{N}_1$ and $\vec{N}_2$ lie in the $(xOz)$ plane and that their sum goes along $(Oz)$. $\varphi$ is the dihedral angle between the planes spanned by $(\vec{N}_1,\vec{N}_2)$ and $(\vec{N}_3,\vec{N}_4)$ }
	\label{fig:interal_parallelogram_and_angle}
\end{figure}

Tetrahedra are the classical counterpart of the basic quanta of geometry carried by spin network states in loop quantum gravity. They are thought as dual to 4-valent nodes of spin networks \cite{Barbieri:1997ks,Baez:1999tk}: each edge attached to the 4-valent nodes is imagined as transverse to a triangle, whose area is given in Planck units by the spin carried by the edge, and those triangles are assembled to form a tetrahedron. Thus the natural parametrization of tetrahedra in the context of loop quantum gravity is to fix the areas of the four triangles, that is the norms $N_{i}=|\vN_{i}|$ of the normal vectors, and explore the remaining two-dimensional space of tetrahedra. This space is topologically a 2-sphere and is conveniently parametrized\footnotemark{} by the norm  $|\vN_{12}|=|\vN_{1}+ \vN_{2}|$ and the dihedral angle $\vphi$ between the planes spanned by $(\vec{N}_1,\vec{N}_2)$ and $(\vec{N}_3,\vec{N}_4)$, as illustrated on fig.\ref{fig:interal_parallelogram_and_angle},
\be
\cos \varphi = \frac{(\vN_1 \wedge \vN_2).(\vN_3 \wedge \vN_4)}{|\vN_1 \wedge \vN_2| |\vN_3 \wedge \vN_4|}.
\ee
\footnotetext{We consider the internal parallelograms: choosing two pairs of opposite edges, we consider the inscribed parallelogram whose vertices are the midpoints of those four edges. This provides three parallelograms whose area are given by $N_{12}=|\vN_{1}+ \vN_{2}|=N_{34}$, $N_{13}=|\vN_{1}+ \vN_{3}|=N_{24}$ and $N_{14}=|\vN_{1}+ \vN_{4}|=N_{23}$. They satisfy the relation:
\be
N_{12}^{2}+N_{13}^{2}+N_{14}^{2}=\sum_{i}N_{i}^{2}\,.
\nn
\ee
Exploring this 2-sphere of internal parallelogram areas satisfying this condition allows to cover the space of tetrahedra with fixed triangle areas $N_{i}$.
}
These two variables form a canonical pair of the Kapovich-Millson phase space for the tetrahedron \cite{kapovich1996,Baez:1999tk,Bianchi:2012wb}. 

\medskip

The set of six parameters $(N_1,N_2,N_3,N_4,N_{12},\varphi)$ provides a full parametrization of the tetrahedron. We would like to explore how the (dual) quadrupole moment, defined in terms of the normal vectors $\vN_{i}$ reflects the shape of the tetrahedra and quantifies how far it is from the equilateral case which we intuitively identify as the most spherical case. Actually, the equilateral tetrahedron corresponds to the case $(N,N,N,N,\frac{2N}{\sqrt{3}},\frac{\pi}{2})$ and the quadrupole moment $T^{ab}$ of an equilateral tetrahedron is exactly proportional to the identity, $T^{ab}_{\textrm{equi}}\propto\id$, thus justifying to identify it as the spherically symmetric case for $P=4$.

We can proceed by two different methods. First, if we want to reconstruct the quadrupole moment matrix $T^{ab}$, which is not invariant under 3d rotations, from the rotation-invariant data $(N_1,N_2,N_3,N_4,N_{12},\varphi)$, we need to choose a frame in order to gauge-fix the invariance under 3d rotations and reconstruct the tetrahedron in that frame. For instance, we can fix the vectors $\vec{N}_1$ and  $\vec{N}_2$ in the the $(xOz)$  plane with $\vec{N}_{12}$ along the $z$-axis, as on fig.\ref{fig:interal_parallelogram_and_angle}.  This entirely fixes the possible 3d rotations of the tetrahedron and it is possible to fully reconstruct the normal vectors $\vN_{i}$ from the four triangle areas $N_{i}$ and the canonical pair  $(N_{12},\varphi)$. Details are given in appendix \ref{ap:tetrahedron_angle_area_param}. This allows to compute the quadrupole matrix $T^{ab}$ in terms of the parameters $(N_1,N_2,N_3,N_4,N_{12},\varphi)$. 

The second method is to focus on the $\SO(3)$-invariant information encoded in the quadrupole matrix $T^{ab}$, that is its eigenvalues, or equivalently its traces $\tr\,T$, $\tr\,T^{2}$, $\tr\,T^{3}$. These can be entirely expressed in terms of the triangle areas $N_{i}$ and the scalar products between normals which can be expressed in terms of the parameters $(N_1,N_2,N_3,N_4,N_{12},\varphi)$.
To start with, two scalar products do not depend on the dihedral angle $\vphi$ but only on the parallelogram area $N_{12}$:
\be
\vN_{1}\cdot\vN_{2}=
\f12\big{[}
N_{12}^{2}-N_{1}^{2}-N_{2}^{2}
\big{]}
\,,\qquad
\vN_{3}\cdot\vN_{4}=
\f12\big{[}
N_{12}^{2}-N_{3}^{2}-N_{4}^{2}
\big{]}
\,.
\nn
\ee
The other scalar products can be extracted from the definition of $\vphi$ from the scalar product between $(\vN_{1}\w\vN_{2})$ and $(\vN_{3}\w\vN_{4})$:
\be
\cC=(\vN_{1}\w\vN_{2})\cdot(\vN_{3}\w\vN_{4})
=
\cos\vphi\,{ \sqrt{N_1^2 N_2^2 - (\vN_{1}\cdot\vN_{2})^2}\sqrt{ N_3^2 N_4^2 - (\vN_{3}\cdot\vN_{4})^2}}
\,,
\ee
\be
\vec{N}_1\cdot\vec{N}_3
=
-\f1{N_{12}^2}\,
\left[
\,\cC-\f14(N_{1}^2-N_{2}^2+N_{12}^2)(N_{3}^2-N_{4}^2+N_{12}^2)
\right]
\,,
\ee
\be
\left|
\begin{array}{lcl}
\vN_{2}\cdot\vN_{4}
&=&
\vec{N}_1\cdot\vec{N}_3+\f12(N_{1}^2+N_{3}^2-N_{2}^2-N_{4}^2) \\
\vN_{1}\cdot\vN_{4}
&=&
-\vec{N}_1\cdot\vec{N}_3-\f12(N_{1}^2-N_{2}^2+N_{12}^2) \\
\vN_{2}\cdot\vN_{3}
&=&
-\vec{N}_1\cdot\vec{N}_3-\f12(N_{3}^2-N_{4}^2+N_{12}^2)
\end{array}
\right.
\nn
\ee
%
As we will see later, these formulas for the scalar products drastically simplify in the iso-area case, when the triangle areas are all equal.

\medskip

So we focus on how the three  $\SO(3)$-invariant observables, $\tr\,T$, $\tr\,T^{2}$, $\tr\,T^{3}$, although replacing $\tr\,T^{3}$ by $\det\,T$, describe the shape of the tetrahedron,
\be
T^{ab} = \sum_{i=1}^{4} \frac{1}{N_i} N_i^{a} N_i^{b}
\,,\quad
\tr\,T=\sum_{i}N_{i}=\cA
\,,\quad
\tr\,T^{2}=\sum_{i,j}\f1{N_{i}N_{j}}\big{(}\vN_{i}\cdot\vN_{j}\big{)}^{2}
\, , \quad
\det\,T = \left(\frac{9}{2}\right)^{2} \frac{\cA}{N_1 N_2 N_3 N_4} V^{4}
\,,
\ee
where $V$ is the volume of the tetrahedron.
Let us keep the triangle areas $N_{i}$ fixed. Then $\tr\,T$ is just the area of the tetrahedron and is also held fixed. The shape of the tetrahedron is characterized by the two remaining observables, $\tr\,T^{2}$ and $\det \,T$, which should carry the same information as the canonical pair $(N_{12},\vphi)$.

The determinant $\det\,T$ measures the volume $V$ of the tetrahedron, while $\tr\,T^2$ should thus measure the non-spherical shape  (or anisotropy) of the tetrahedron.
It is more convenient to replace $\tr\,T^{2}$ by the trace of the square of the traceless quadrupole matrix,
\begin{equation}
\tT=T-\f13 \tr \,T\,\id
\,,\qquad
\tr\,\tilde{T}^2 = \tr\,T^2 - \f13 \cA^2
\,.
\end{equation}
$\tilde{T}$ has the nice property to vanish if and only if the tetrahedron is equilateral.

\medskip

\begin{figure}[!htb]
	\includegraphics[height=35mm]{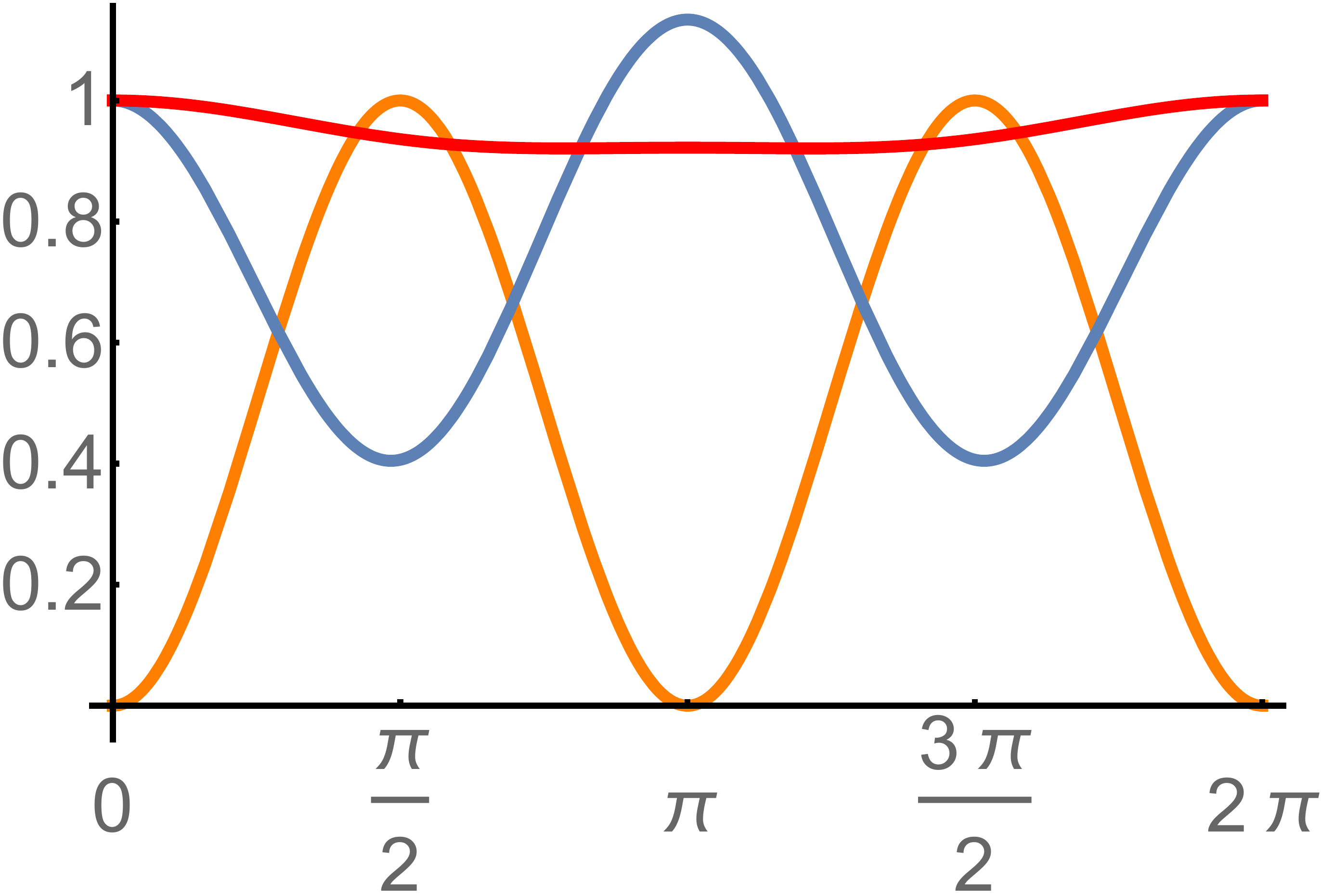}
	\hspace*{20mm}
	\includegraphics[height=35mm]{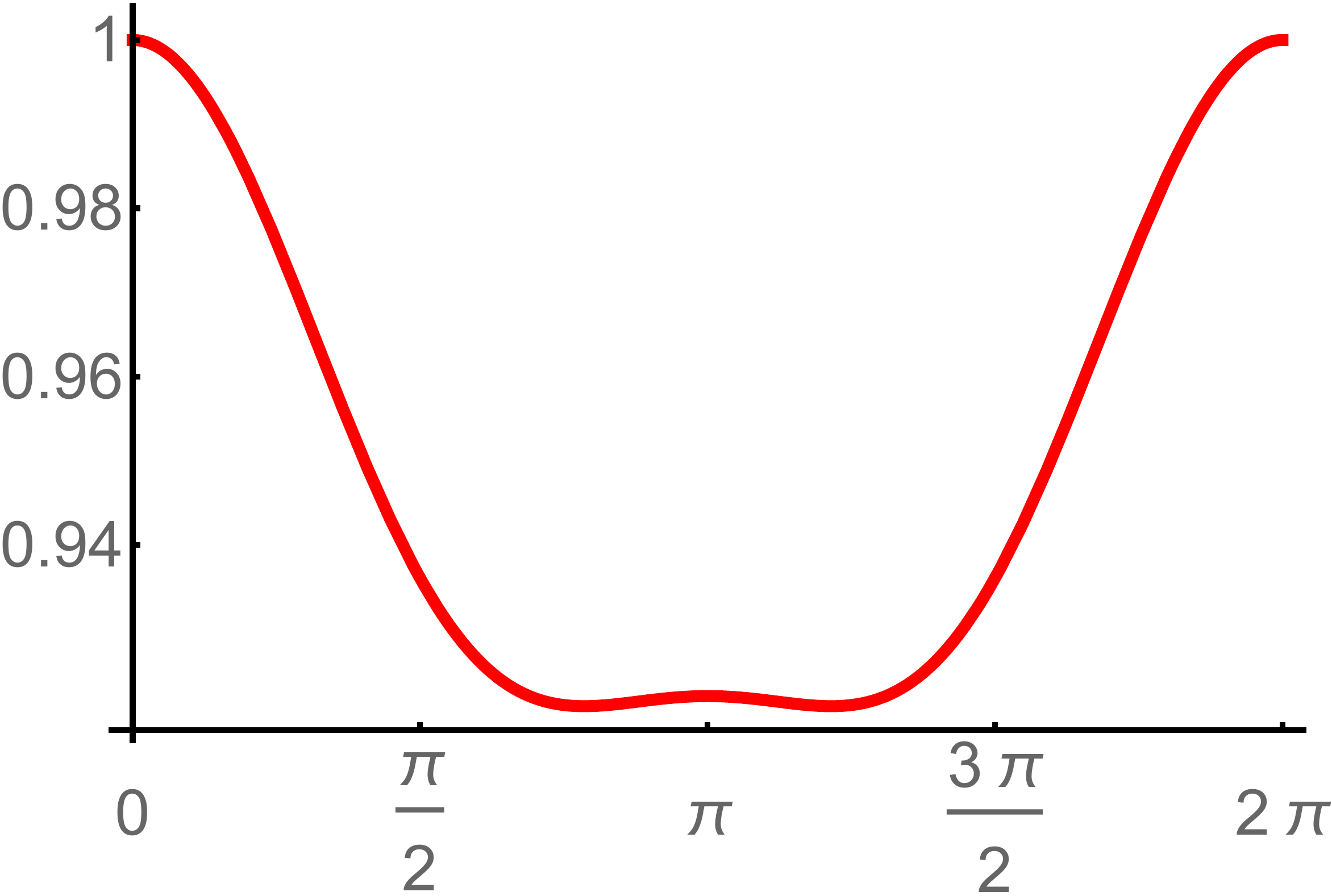}
	\caption{
Plots of $\frac{\det\,T}{\det\,T|_{\varphi=\frac{\pi}{2}}}$ (orange curve vanishing at $\vphi=0$) and $\frac{\tr\,\tilde{T}^2}{\tr\,\tilde{T}^{2}|_{\varphi=0}}$ (blue and red curves starting at 1 at $\vphi=0$).
The curve for the volume observable $\det\,T$ is normalized by its maximal value $\det\,T|_{\varphi=\frac{\pi}{2}}$ reached at $\vphi=\f\pi2$ and does not depend on the area data $(N_{1},N_{2},N_{3},N_{4},N_{12})$.
The shape observable $\tr\,\tilde{T}^2$ is normalized by its initial value for the degenerate tetrahedron with $\vphi=0$ and does depend on the area data. We plotted the blue curve for $(N_{1},N_{2},N_{3},N_{4},N_{12})=(7,9.4,5.4,5.1,4)$ and the red curve -the flatter one- for values  $(12,4,7,3,8.2)$. We provide a zoom of the red curve on the right hand side to show that $\tr\,\tilde{T}^2$  reaches its minimal value for an angle $\vphi_{min}$ which depends on the area data and then reaches another local maximum at $\vphi=\pi$.}
	\label{fig:plot_detT_TrT2}
\end{figure}

We have plotted the evolution of $\det\,T$ and $\tr\,\tilde{T}^2$ in terms of the dihedral angle $\varphi$ running  from $0$ to $2\pi$ while keeping the area data $N_1$, $N_2$, $N_3$, $N_4$ and $N_{12}$ fixed. On the fig.\ref{fig:plot_detT_TrT2}, we can see how these two observables oscillate with $\varphi$ reaching respectively their minima and maxima at $0$, $\f\pi2$ or $\pi$.
First of all, there are two possible orientations for tetrahedra, corresponding to a tetrahedron and its mirror image. These two tetrahedra are not related by a 3d rotation, but have the same area, volume and shape. This leads to a symmetry under $\varphi \rightarrow 2 \pi - \varphi$.  As we can see on  fig.\ref{fig:plot_detT_TrT2}, the curves for $\tr\,\tilde{T}^2$ and $\det\,T$ are indeed symmetric with respect to $\pi$.

Two important remarks need to be made about the volume observable $\det\,T$. It is also symmetric under reflection with respect to $\frac{\pi}{2}$. And, once properly normalized by its maximal value reached at $\vphi=\f\pi2$, it does not depend on the area data but only on the dihedral angle $\vphi$. Actually the exact formula for the volume square is\footnotemark{} (see also e.g. \cite{Bianchi:2012wb}):
\be
\label{Vol}
V^{2}=\f29\,\big{|}
\vN_{1}\cdot(\vN_{2}\w\vN_{3})
\big{|}
\,,\quad
\vN_{1}\cdot(\vN_{2}\w\vN_{3})
=
\f{\Delta\bar{\Delta}}{4N_{12}}\,\sin\vphi
\quad\textrm{with}\,\,
\left|
\begin{array}{lcl}
\Delta&=& \sqrt{(N_{1}+N_{2})^{2}-N_{12}^{2}}\,\sqrt{N_{12}^{2}-(N_{1}-N_{2})^{2}}
\\
\bar{\Delta}&=& \sqrt{(N_{3}+N_{4})^{2}-N_{12}^{2}}\,\sqrt{N_{12}^{2}-(N_{3}-N_{4})^{2}}
\end{array}
\right.
\ee
\footnotetext{
The volume formula can be easily derived from the double cross-product formula:
\be
(\vN_{1}\w\vN_{2})\w(\vN_{3}\w\vN_{4})
=
(\vN_{1}\cdot(\vN_{2}\w\vN_{3}))\,\vN_{12}
\,,
\nn
\ee
where we have taken into account the closure constraint, $\vN_{4}=-(\vN_{1}+\vN_{2}+\vN_{3})$. This implies that:
\be
\vN_{1}\cdot(\vN_{2}\w\vN_{3})
=
\f{|\vN_{1}\w\vN_{2}|\,|\vN_{3}\w\vN_{4}|\,\sin\vphi}{N_{12}}
\,,
\nn
\ee
and we can write the norms $|\vN_{1}\w\vN_{2}|$ and $|\vN_{3}\w\vN_{4}|$ using the Heron formula for triangle areas.
}

On the other hand, $\vphi=\f \pi2$ is not a special point for the shape observable $\tr\tT^{2}$ and it is not invariant under the symmetry $\vphi\rightarrow\pi-\vphi$. The quadrupole moment $\tr\tT^{2}$ reaches its two maxima at $\vphi=0$ and $\vphi=\pi$, although its two values are generally not  equal, and reaches its minimum at an angle $\vphi_{min}$ which depends on the area data (and is not generally $\f \pi2$). This minimum corresponds to the tetrahedron closest to the equilateral case for fixed area data and is actually located at $\vphi_{min}=\f\pi2$ for iso-area tetrahedra, as shown below.

At the end of the day, at fixed triangle areas, it is possible to switch the canonical area-angle pair $(N_{12},\vphi)$ for the pair of geometrical observables $(\tr\,\tT^{2}\,,\, \det\,T)$ encoding the volume and the shape of the tetrahedron. This holds up to taking the mirror image of tetrahedra. Moreover, we have plotted the Jacobian of this change of variable in  fig.\ref{fig:jacobian}  in appendix \ref{ap:jacobian} showing that it only has isolated zeroes.

\subsection{The iso-area tetrahedron}

\begin{figure}[h!]
	\begin{center}
		\includegraphics[height=40mm]{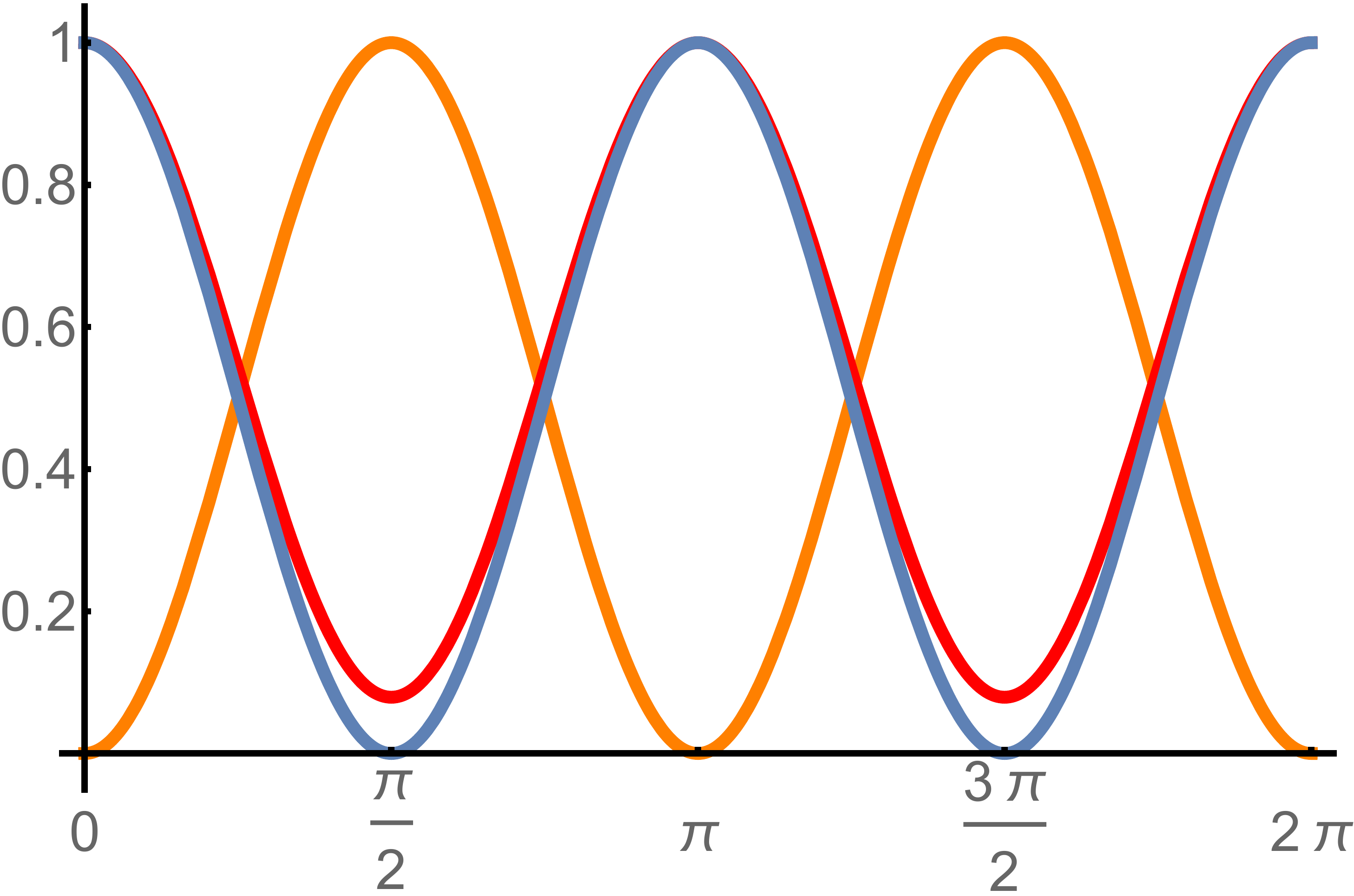}
	\end{center}
	\caption{
Plots of ${\det\,T}$ (orange) and ${\tr\,T^2}$ (blue and red) normalized by their maximal value reached respectively at $\vphi=\f\pi2$ and $\vphi=0$.
The blue curve corresponds to the area data $N_{12}=\frac{2N}{\sqrt{3}}$, i.e the case where we obtain an equilateral tetrahedron for $\varphi = \frac{\pi}{2}$. The red curve corresponds to the area data $N_{12}=\f89N$.}
	\label{fig:plot_detT_trT2_iso}
\end{figure}

In this section, we specialize to the simpler case of  iso-area tetrahedra, i.e. the case where the four areas of the triangles are equal $N_{i}=N$ for all $i=1..4$. Now both the volume observable $\det\,T$ and the shape observable $\tr\,\tT^{2}$ are symmetric under $\vphi\rightarrow \pi-\vphi$. As illustrated on fig.\ref{fig:plot_detT_trT2_iso}, they vary exactly out of phase: $\det\,T$ is minimal for the degenerate tetrahedra at $\vphi=0$ and $\vphi=\pi$ and maximal at $\vphi=\f\pi2$, while $\tr\,\tT^{2}$ is maximal for the degenerate cases at $\vphi=0$ and $\vphi=\pi$ and minimal for the configuration closest to the equilateral tetrahedron at  $\vphi_{min}=\f\pi2$.

We can also provide explicit formulas. We start with reconstructing the normals in terms of $N$, $N_{12}$ and $\vphi$, when fixing $\vN_{1}$ and $\vN_{2}$ in the $(xOz)$ plane as on fig.\ref{fig:interal_parallelogram_and_angle}.  Details are given in appendix \ref{ap:tetrahedron_angle_area_param}. This allows to compute the (dual) quadrupole matrix:
\begin{equation}
T^{ab} =2 N \begin{pmatrix}
s^{2}_\frac{\alpha}{2} (1+c^{2}_{\varphi}) & s^{2}_{\frac{\alpha}{2}} c_{\varphi} s_{\varphi} & 0 \\
s^{2}_{\frac{\alpha}{2}} c_{\varphi} s_{\varphi} & s^{2}_{\frac{\alpha}{2}} s^{2}_{\varphi} & 0 \\
0 & 0 & 2 c_{\frac{\alpha}{2}}^{2}
\end{pmatrix}
\,,\qquad
c_{\f\alpha2} =\f{N_{12}}{2N}\,.
\end{equation}
where we have used the more compact notation $c$ and $s$ for the cosines and sines.
$\vphi$ is the dihedral angle between the $(\vN_{1},\vN_{2})$ plane and the $(\vN_{3},\vN_{4})$ plane, while $\alpha$ is the angle between the two vectors $\vN_{1}$ and $\vN_{2}$.
We diagonalize $T^{ab}$ to determine its eigenvalues:
\begin{equation}
O\, T \, {}^{t}O = 4N \begin{pmatrix}
s_{\frac{\alpha}{2}}^{2} c^{2}_{\frac{\varphi}{2}} & 0 & 0 \\
0 & s_{\frac{\alpha}{2}}^{2} s^{2}_{\frac{\varphi}{2}} & 0 \\
0 & 0 & c_{\frac{\alpha}{2}}^{2}
\end{pmatrix}
\quad \text{with} \quad
O = \begin{pmatrix}
c_{\frac{\varphi}{2}} & s_{\frac{\varphi}{2}} & 0 \\
-s_{\frac{\varphi}{2}} & c_{\frac{\varphi}{2}} & 0 \\
0 & 0 & 1
\end{pmatrix},
\end{equation} 	
which entirely determines the  ellipsoid approximating the tetrahedron. It is interesting to note that the three axis of the ellipsoid correspond to the three normals to the internal parallelograms, $\vN_{12}$, $\vN_{13}$ and $\vN_{14}$, which form an orthogonal basis.

We compute the volume observable:
\be
\det\,T
=
4^{2}N^{3}s^{4}_{\f\alpha 2}c^{2}_{\f\alpha 2}s^{2}_{\vphi}
=
\f{N_{12}^{2}(4N^{2}-N_{12}^{2})^{2}}{4N^{3}}\,\sin^{2}\vphi
\,,
\ee
which equates the general formula \eqref{Vol} in the iso-area case, and we compute the shape observable:
\be
\tr\,\tT^{2}
=
(4N)^{2}\left[
\f12s^{4}_{\f\alpha2}(1+\cos^{2}\vphi)+c^{4}_{\f\alpha2}-\f13
\right]
\,,
\ee
which does vanish as expected  for the equilateral tetrahedron, when $\cos\f\alpha2=\f19$ and $\vphi=\f\pi2$.

\section{The Shape of Quantum Polyhedra}


Beyond the simplest case of a tetrahedron corresponding to a 4-valent node of a spin network, intertwiners living at a $P$-valent node are interpreted as quantum polyhedra with $P$ faces. This has been confirmed by the construction of suitable coherent intertwiner states, maximally peaked on classical polyhedra and thus providing the notion of semi-classical polyhedra \cite{Livine:2007vk,Freidel:2009nu,Freidel:2010tt, Bianchi:2010gc,Livine:2013tsa}. This has allowed the concrete geometrical interpretation of spin network states as semi-classical discrete geometries in the twisted geometry framework \cite{Freidel:2010aq}.

The now-called Livine-Speziale (LS) intertwiners introduced in \cite{Livine:2007vk,Livine:2007ya} are group-averaged tensor products of $\SU(2)$ coherent states. Their interpretation as semi-classical polyhedra is as follows. Each $\SU(2)$ coherent state encodes the quantum equivalent of a 3-vector, interpreted as the normal vector of a face. The group-averaging over $\SU(2)$ has a double role, on the one hand, leading to states invariant under 3d rotations, and on the other hand, imposing the closure constraints on the normal vectors implying the existence of a unique classical convex polyhedron corresponding to this configuration of normal vectors. 

Nevertheless, the seminal analysis in \cite{Livine:2007vk} left one open question. Looking at the large spin behavior of the coherent intertwiners by a saddle point analysis, it was shown that the existence of a stationary point imposed the closure constraint, while the Hessian matrix $H$ controlling the spread of the intertwiners was given explicitly in terms of the normal vectors, but without any geometrical interpretation. Here we address this issue and we can identify the coherent intertwiner Hessian $H$ as the quadrupole moment $T$. Thus the quadrupole moment  also controls the shape of quantum polyhedra. 

\medskip

Let us show how the coherent intertwiner Hessian leads back to the quadrupole moment. We start by defining $\SU(2)$ coherent states of spin $j$ labeled by a spinor $z\in\C^{2}$:
\be
|j,z\ra=\sqrt{(2j)!}\,\sum_{m}\f{(z^{0})^{j+m}(z^{1})^{j-m}}
{\sqrt{(j+m)!(j-m)!}}
\,|j,m\ra
\,,
\qquad\textrm{with}\quad
z=\mat{c}{z^{0}\\ z^{1}}\,\in\C^{2}
\,,
\ee
where the $|j,m\ra$ are the usual basis states of the irreducible $\SU(2)$-representation of spin $j$ labeled by the magnetic index $m$.
These are coherent states \`a la Perelomov \cite{perelomov1972} and were re-introduced in the context of spin networks and loop quantum gravity in \cite{Livine:2007vk,Freidel:2007py,Freidel:2010tt,Borja:2010rc}.
For the fundamental representation, this definition is tautological and $|\f12,z\ra\,=\,|z\ra$ as a complex vector in $\C^{2}$. For arbitrary spin $j$, these states are homogeneous of degree $2j$, i.e. $|j,\lambda z\ra=\lambda^{2j}\,|j,z\ra$ for arbitrary rescaling $\lambda\in\C$.
Their norm and scalar products are immediate to compute:
\be
\la j,z|j,z\ra= \la z|z\ra^{2j}
\,,\qquad
\la j,w|j,z\ra=\la w|z\ra^{2j}
\,.
\ee
The first key property is that they transform coherently under the $\SU(2)$ action:
\be
D^{j}(g)\,|j,z\ra
\,=\,
|j,g\,z\ra
\,,
\ee
where group elements $g\in\SU(2)$ act on spinors $z\in\C^{2}$ as 2$\times$2 matrices in the fundamental representation $j=\f12$.
Their second key property is that they are semi-classical states peaked on classical vectors of discrete length $j$. Consider normalized spinors, $\la z|z\ra=1$, thus normalized coherent states, we compute the expectation values of the $\su(2)$ generators $J^{a=1,2,3}$ on those states:
\be
{\la j,z| \vJ|j,z\ra}
\,=\,
j\,\la z| \vsigma|z\ra
\,=\,
j\,\hat{n}
\qquad\textrm{with}\quad
|\hat{n}|^{2}=|\la z| \vsigma|z\ra|^{2}=|\la z|z\ra|^{2}=1
\,,\quad
\hat{n}\in\cS_{2}\,,
\ee
where the $\sigma^{a}$ are the Pauli matrices.
We will call $\vN=j\hn$ the expectation value vector, which has the quantized length $|\vN|=j$.

Now we consider $P$ such coherent states, $|j_{i},z_{i}\ra$ with $i=1..P$, and we define the LS coherent intertwiner as the group-averaged tensor product of those states,
\be
||\{j_{i},z_{i}\}\ra
\,=\,
\int_{\SU(2)}\dd g\,
\bigotimes_{i=1}^{P}D^{j_{i}}(g)\,|j_{i},z_{i}\ra
\,.
\ee
The group averaging ensures to produce $\SU(2)$-invariant states, i.e. intertwiners in the tensor product $j_{1}\otimes..\otimes j_{P}$. It was showed  in \cite{Livine:2007vk} that these states are peaked on closed configurations, i.e. satisfying the closure constraints $\sum_{i}\vN_{i}=0$. This was achieved through the semi-classical analysis at large spins using the saddle point approximation for the group integral\footnotemark{}. 
\footnotetext{
One can also find the explicit  expansion in \cite{Freidel:2010tt,Bonzom:2012bn} in terms of products of scalar products between spinors $z_{i}$. 
}
But the question of geometrical meaning of Hessian of the saddle point approximation was left open.
So let us look into the semi-classical  behavior of the LS coherent intertwiner norm at large spins. This is the archetype computation for large spin asymptotics of spinfoam path integral amplitudes using semi-classical boundary spin network states, see e.g. \cite{Barrett:2010ex, Freidel:2013fia}.

The norm of a LS coherent intertwiner is written as an integral over $\SU(2)$ of a complex action depending on the group element $g$ and the spinors $z_{i}$:
\be
\la \{j_{i},z_{i}\}||\{j_{i},z_{i}\}\ra
\,=\,
\int_{\SU(2)} \dd g\,
\prod_{i=1}^{P}
\la z_{i}|g|z_{i}\ra^{2j_{i}}
=
\int \dd g\,
e^{S[g,\{z_{i}\}]}
\quad\textrm{with}\quad
S\equiv\sum_{i}^{P}2j_{i} \ln \la z_{i}|g|z_{i}\ra
\,.
\ee
Focusing on the large spin behavior (when all the spins are homogeneously rescaled $j_{i}\rightarrow k j_{i}$ with $k$ sent to infinity), the behavior of this integral is controlled by its stationary points in $g$.
Since each factor $\la z_{i}|g|z_{i}\ra$ is bounded in modulus by 1 and only reach this maximal modulus when at the identity $g=\id$, the maximum of the  integrand is clearly reached at $g=\id$, at least in modulus. We nevertheless have to check that it is actually a fixed point in phase. Computing the first derivative of $S[g,\{z_{i}\}]$ with respect to $g$ and evaluating at the identity actually gives the dipole moment:
\be
\pp^{a}S[g,\{z_{i}\}]\Big{|}_{g=\id}
\,=\,
\sum_{i} 2j_{i}\f{\la z_{i}|g\sigma^{a}|z_{i}\ra }{\la z_{i}|g|z_{i}\ra }\Big{|}_{g=\id}
\,=\,
\sum_{i} 2j_{i}\la z_{i}|\sigma^{a}|z_{i}\ra 
\,=\,
\sum_{i} 2j_{i}\hn_{i}^{a}
\,=\,
2\sum_{i}N_{i}^{a}
\,.
\ee
Thus the existence of a fixed point at $g=\id$ implies the closure constraint, $\sum_{i}\vN_{i}=0$. When the closure constraint is relaxed and we have a non-trivial closure defect  $\sum_{i}\vN_{i}\ne0$, the LS intertwiner norm is exponentially suppressed and its decay is given by the norm square of the closure defect \cite{Livine:2007vk}.

Next, assuming the closure constraint, we compute the Hessian at $g=\id$ as the second derivative of the action and get:
\be
\pp^{b}\pp^{a}S[g,\{z_{i}\}]\Big{|}_{g=\id}
\,=\,
\sum_{i}2j_{i}\big{(}\delta^{ab}-\hn_{i}^{a}\hn_{i}^{b}\big{)}
\,=\,
2\left(
\cA\delta^{ab}
-\sum_{i}\f1{N_{i}}N_{i}^{a}N_{i}^{b}
\right)
\,=\,
2\left(
\cA\delta^{ab}
-T^{ab}
\right)
\,,
\ee
where we recognize both the total area $\cA=\tr \,T$ and the quadrupole moment matrix. In the original work \cite{Livine:2007vk} on coherent intertwiners, the inverse norm factor  $\f1{N_{i}}$ was puzzling and was not given a geometrical interpretation, while it appears in our context to ensure the re-parametrization invariance of the multipole moments in the continuum as explained above in section \ref{multipole}. At the end of the day, the quadrupole moment entirely controls the large spin behavior of the LS intertwiners and we have in particular for their norm (up to factors $\pi$ and 2), as long as the closure constraint is satisfied:
\be
\la \{j_{i},z_{i}\}||\{j_{i},z_{i}\}\ra
\propto
\f1{\sqrt{\det (\cA\id
-T)}}
\,,
\ee
which reproduces the expected scaling in $j^{{-\f32}}$. Finally the determinant $\det (\cA\id-T)$ can be easily expanded and computed in terms of the power traces $\tr\, T^{k}$ for $1\le k \le 3$.

\section{A Quadrupole Operator in Loop Quantum Gravity}
\label{operator}

Finally, to probe the deep quantum regime after the previous investigation of the classical and semi-classical frameworks, we now turn to the definition of a quadrupole moment operator acting on intertwiners. If we only seek operators acting on intertwiners, i.e. on $\SU(2)$-invariant states, we should focus on the $\SO(3)$-invariant components of the quadrupole tensor, that its three eigenvalues or equivalently its power traces $\tr\,T$, $\tr\,T^2$ and $\tr T^3$. Nevertheless, it is much more interesting and complete to quantize all the matrix elements of the quadrupole tensor $T^{ab}$, and construct its power traces as polynomials in these, especially in the prospect of building operators correlating the quadrupole tensors between different vertices of a spin network or of studying general quantum surface states in loop quantum gravity (see e.g. \cite{Feller:2017ejs}).

Considering the classical expression of the quadrupole moment, 
\be
T_{ab}=\sum_{i}\f1{N_{i}}N_{i}^aN_{i}^b\,,
\nn
\ee
we expect two quantization issues: the inverse norm factor $N_{i}^{-1}$ and the ordering of the two terms $N_{i}^a$ and $N_{i}^b$. We will fix the regularization of the inverse norm factor by imposing to maintain at the quantum level the fact that the quadrupole trace is the area, $\tr\,T=\cA$. As for the ordering ambiguity, it will most likely be fixed by imposing to maintain certain relevant Poisson brackets as exact commutators at the quantum level. However, the Poisson brackets of the matrix elements $T_{ab}$'s or the power tracers $\tr\,T^k$ involves higher multipole moments and studying the whole Lie algebra of dual multipole moments is out of the scope of the present work, even though it definitely remains a point to be studied further. Here, we will introduce two natural orderings, the straightforward one and the symmetric one, and we will discuss their properties.

The classical Poisson bracket on the normal vectors is:
\be
\{N_{i}^a,N_{j}^b\}
\,=\,
\delta_{ij}\,\eps^{abc}N_{i}^c.
\ee
Provided with the closure constraint $\sum_{i}\vN_{i}=0$, this gives the Kapovich-Milson phase space for polyhedra \cite{kapovich1996,Livine:2013tsa}. Upon quantization, it leads to $P$ copies of the $\su(2)$ Lie algebra, with the vector components $N_{i}^a$ becoming the $\su(2)$ generators $J_{i}^a$ satisfying the usual commutators:
\be
\hat{N}_{i}^a=J_{i}^a
\,,\qquad
[J_{i}^a,J_{j}^b]=\,i\,\delta_{ij}\eps^{abc}J_{i}^c\,.
\ee
The Hilbert space consists of $P$ copies of the Hilbert space for a single face. Each face can carry an arbitrary spin $j_{i}$, so that the face Hilbert space consists in the direct sum of all spins:
\be
\cH_{P}=\cH^{\otimes P}\,,
\qquad
\cH=\bigoplus_{j\in\N/2}V^j
\,.
\ee
The space $V^j$ carries the spin-$j$ representation of $\SU(2)$. It is $(2j+1)$-dimensional and its usual basis $|j,m\ra$ is labeled by the magnetic index $m$ running by integer step from $-j$ to $+j$.

Here, we follow the operator ordering of the spinorial formulation based on the Schwinger realization of the $\su(2)$ Lie algebra as a pair of harmonic oscillators \cite{Girelli:2005ii,Freidel:2009ck,Borja:2010rc,Livine:2011gp,Livine:2013tsa}. This leads to the equidistant area spectrum and the quantization of the normal vector norm $N_{i}$ as the simple spin operator:
\be
\hat{N}_{i}\,|j_{i},m_{i}\ra
\,=\,
j_{i}
\,|j_{i},m_{i}\ra\,.
\ee
First, this means that the total area of the quantum surface is $\hat{\cA}=\sum_{i}j_{i}$. 
Second, the operator $N_{i}$ has a zero eigenvalue and is not invertible. Therefore we have to regularize it in order to define the inverse norm factors of the quadrupole operators.

Since all the operators $J_{i}^a$ don't change the spins $j_{i}$ and commute with $\hat{N}_{i}$, we will write $j_{i}$ for $\hat{N}_{i}$ for simplicity's sake and work at fixed spins on all the faces unless stated otherwise. The straightforward ordering we propose for the $T$-operators is:
\be
\hat{T}^{ab}=\sum_{i}\f1{j_{i}+1}J_{i}^aJ_{i}^b
\,,
\ee
while the symmetric ordering uses the anti-commutator of the two $\su(2)$-generators:
\be
\hat{\cT}^{ab}
=\sum_{i}\f1{j_{i}+1}\,\{J_{i}^a,J_{i}^b\}
=\f12\sum_{i}\f1{j_{i}+1}\,(J_{i}^aJ_{i}^b+J_{i}^bJ_{i}^a)
\,.
\ee
The shifted inverse spin $(j+1)^{-1}$, instead of the ill-defined $j^{-1}$, ensures that both orderings properly implements the classical relation $\tr\,T=\cA$. Indeed:
\be
\sum_{a}T^{aa}=\sum_{a}\cT^{aa}
=\sum_{i}\f{\vJ_{i}{}^2}{j_{i}+1}
=\sum_{i}\f{j_{i}(j_{i}+1)}{j_{i}+1}
=\sum_{i}j_{i}
=
\hat{\cA}
\,.
\ee
From the operators $T^{ab}$ or $\cT^{ab}$, we can define the higher power trace operators. For instance, the quantization of $\tr\,T^2$ gives:
\be
\hT^{ab}\hT^{ba}=
\sum_{i,j}\f{(\vJ_{i}\cdot\vJ_{j})^{2}}{(j_{i}+1)(j_{j}+1)}
+
\sum_{i\ne j}\f{(\vJ_{i}\cdot\vJ_{j})}{(j_{i}+1)(j_{j}+1)}
\,,
\ee
\be
\hcT^{ab}\hcT^{ba}=
\sum_{i,j}\f{(\vJ_{i}\cdot\vJ_{j})^{2}}{(j_{i}+1)(j_{j}+1)}
-
\f12\sum_{i}\f{\vJ_{i}{}^{2}}{(j_{i}+1)^{2}}
+
\f12\sum_{i\ne j}\f{(\vJ_{i}\cdot\vJ_{j})}{(j_{i}+1)(j_{j}+1)}
\,.
\ee
One can also easily compute the quantum operator for $\tr\,T^3$ for both choices of ordering, which will similarly contain subdominant contributions coming from the operator ordering. This set of three operators, $\widehat{\tr\,T}$, $\widehat{\tr\,T^2}$ and $\widehat{\tr\,T^3}$, are $\SU(2)$-invariant and act on intertwiners. Although the first one, $\widehat{\tr\,T}$, is trivial in the sense that it  simply gives the total area as the sum of spins, the other two operators, $\widehat{\tr\,T^2}$ and $\widehat{\tr\,T^3}$, are new non-trivial operators probing the shape of intertwiners. It would be very interesting to investigate further if they can be diagonalized. Their spectrum would reveal how shape is quantized in loop quantum gravity.

\section*{Conclusion \& Outlook}

%
%

\medskip

In this work, we have focused on the multipole moments of the normal vector field to an embedded surface. At first, we called them ``dual multipole'' to distinguish them for the usual multipole moments defined as polynomial averages of the position. But since we only have access to the normal vector to surfaces in the standard framework of loop quantum gravity, we dropped the adjective ``dual'' and defined the multipole moments for discrete surfaces (polyhedra) and then quantum surfaces.

Having identified the (dual) monopole as the surface area and the (dual) dipole as the closure defect (or closure constraint for closed surfaces embedded in flat 3d space), we have focused on the (dual) quadrupole moment. This provides us with a new of set of basic observables for loop quantum gravity probing the shape of intertwiners and more generally of quantum surfaces on a spin network state. It will be interesting to diagonalize these operators  and investigate their spectrum. 

\smallskip

The simplest way to understand this new quadrupole moment is by considering the ladder of higher polynomial observables in the $\su(2)$ generators $\vJ_{i}$. The monopole, which gives the area of a surface in loop quantum gravity, will be given by the sum over the spins $j_{i}$, defining the $\su(2)$-representation attached to each face. The dipole looks at $\sum_{i} \vJ_{i}$, which we recognize as the closure defect and which constrain to vanish for intertwiners. The closure constraint  $\sum_{i} \vJ_{i}=0$ more generally identifies closed surfaces. It is then natural to consider higher order observables, such as  $\sum_{i} \vJ_{i}\otimes \vJ_{i}$. The unexpected point is that we actually an extra inverse norm factor to ensure the invariance under re-parametrization in the continuum limit, so that the correct quadrupole moment is defined as
\be
T=\sum_{i}\f1{(j_{i}+1)} \vJ_{i}\otimes \vJ_{i}
\,,
\ee
where the shift in the spins in the denominator, $j\rightarrow (j+1)$, is a regularization at the quantum level.
The eigenvalues and eigenvectors of this operator will define the basic {\it quanta of shape} for loop quantum gravity.

Two points are specially enticing about this new quadrupole observable. First, we showed that the quadrupole moment matrix is exactly the Hessian matrix controlling the semi-classical behavior at large spins of LS coherent intertwiners. This fully clarifies the geometrical meaning of those states beyond the leading order. Second, it turns out that the quadrupole $T$ describes the shape of the surface or intertwiner, but also contains information on the  volume bounded by that surface. We discussed the similarity and difference, at the classical level, of the new observable $\det\,T$ with the usual squared-volume observable $U$ used in loop quantum gravity. It would be interesting to push this analysis to the quantum level, and study its spectrum exactly or in a semi-classical approximation \cite{Bianchi:2012wb}. Perhaps, $\det\,T$ will provide a better behaved volume observable (in particular, since it is Hermitian positive and does not depend on the bivector space orientation), useful for the dynamics of loop quantum gravity.

We also would like to mention two interesting threads to follow, in order to better understand the mathematical properties of the quadrupole moment. On the one hand, it is natural to investigate the whole towers of multipole moments, identify their Lie algebra under the loop gravity Poisson bracket and look for a suitable representation as quantum operators acting on intertwiners. Probably, this should be investigated in the more general framework of quantum surface states defined in \cite{Freidel:2016bxd}. On the other hand, it should be very interesting to investigate the action of the basic surface deformation operators on the quadrupole moment, that is look at the commutators of the quadrupole moment observables with the $\U(N)$ operators (and generally $\SO^{{*}}(2N)$ operators) describing the basic deformations of intertwiners as elementary exchange of area quanta \cite{Freidel:2009ck,Freidel:2010tt,Livine:2013tsa,Girelli:2017dbk}. Perhaps, this will also shed light on a possible diagonalization of the quadrupole operators.

\smallskip

Finally, we hope that these  quadrupole moment operators, and more generally the multipole observables we introduced, will provide loop quantum gravity with an useful new tool to probe the shape of intertwiners and excite the deformations, in the objective of modeling and studying shape fluctuations and the propagation of shape correlations towards a clearer pictures of geometrical waves in loop quantum gravity.

%

\appendix

\section{Quadrupole, Dual Quadruple and Surface shape}

We use the same notation as in the main text to give more details on the standard and dual quadrupole moments and provide explicit formulae for ellipsoids.

\subsection{Quadrupole moment and Ellipsoid Approximation for a Closed Surface}
\label{app:ellipsoid} 

We parametrize a two-dimensional surface $\cS$ embedded in the 3d flat Euclidean space $\R^{3}$ with two coordinates $(u,v)$ defining the surface points $\vx(u,v)\in\R^{3}$.
We define the multipole moments of the surface $\cS$ by integrating polynomials of the coordinates.
At the zeroth order, the monopole gives the total surface area:
\be
\cI_{0}
\,=\,
\int_{\cS} \dd A
\,=\,
\int_{\cS} \dd u\dd v\,\sqrt{\det g}
\,=\,
\cA\,.
\ee
The dipole moment takes the mean of the positions and defines the center of mass of the surface:
\be
\cI_{1}^{a}
\,=\,
\int_{\cS} \dd u\dd v\,\sqrt{\det g}\,x^{a}
\,=\,
\vX\,.
\ee
The quadrupole moment is defined as a 3$\times$3 symmetric matrix and probes the basic shape of surface:
\be
\cI_{2}^{ab}
\,=\,
\int_{\cS} \dd u\dd v\,\sqrt{\det g} \,x^{a}x^{b}
\,=\,
t^{ab}\,.
\ee
We also usually  define its traceless component:
\be
\tt^{ab}=t^{ab}-\f13\tr t \,\id\,.
\ee
Higher order polynomials and the corresponding moments will explore the finer structure of the surface.%
Under rigid translations of the whole surface, $\vx\arr \vx+\vv$, the area is of course invariant and the center of mass simply gets translated:
\be
x^{a}\arr x^{a}+v^{a}
\quad\Rightarrow\quad
\left|
\begin{array}{lcl}
\cA&\arr&\cA \\
X&\arr&X+v \\
t&\arr& t + X\otimes v+ v\otimes X+ v\otimes v
\end{array}
\right.
\ee
Thus we can always center the surface on the origin by a translation, which does make the dipole vanish but still leaves us with a non-trivial quadrupole tensor:
\be
x^{a}\arr x^{a}-X^{a}
\quad\Rightarrow\quad
\left|
\begin{array}{lcl}
\cA&\arr&\cA \\
X&\arr&0 \\
t&\arr& t - X\otimes X \,.
\end{array}
\right.
\ee
Under 3d rotations, each space coordinate transforms as expected under $\SO(3)$ transformations. The area is  invariant. Once centered on the origin, the center of mass of the surface is also invariant. The quadrupole tensor transforms covariantly, $t\arr {}^{t}o\,t\,o$ for an arbitrary rotation $o\in\SO(3)$. This allows to diagonalize the quadrupole:
\be
\exists o\in\SO(3)\,,\lambda_{a}\in\R^{3}\,,
\quad\textrm{such that}\quad
t={}^{t}o \mat{ccc}{\lambda_{1} & & \\ &\lambda_{2}& \\ &&\lambda_{3} }o\,.
\ee
The three eigenvalues $\lambda_{a}$ are positive and encode the rotation-invariant data of the quadrupole moment. They describe the shape of the surface. Indeed, taking the ansatz of an ellipsoid, they give the three radii of the ellipsoid while the rotation $o$ gives the change of basis from the original orthonormal frame to the proper frame of the ellipsoid defined by its three principal axis.
The order of the three eigenvalues is not relevant and the same data is contained in the three traces, $\tr\,t$, $\tr\, t^{2}$ and $\tr\, t^{3}$, from which we can reconstruct any symmetric (polynomial) observable in the $\lambda_{a}$'s.

\medskip

{\it Example 1: the  Sphere}

\noindent
Let us start by applying these definitions to the unit sphere centered on the origin. We parametrize it using  the angular variables:
\be
\vx=\mat{c}{\sin\theta\cos\phi \\ \sin\theta\sin\phi \\ \cos\theta}
\,,\quad
\theta\in[0,\pi]
\,,\quad
\phi\in[0,2\pi]
\,,\qquad
\sqrt{\det g}=\sin\theta\,.
\ee
We easily compute the moments as trigonometric integrals:
\be
\cI_{0}=\cA=\int_{0}^{\pi}d\theta\,\int_{0}^{2\pi}d\phi\,\sin\theta=4\pi
\,,\qquad
\cI_{1}=X=0
\,,\qquad
\cI_{2}=t=\f{4\pi}{3}\,\id
\ee
The quadrupole moment is proportional to the identity, thus its traceless component vanishes, $\tt=0$. This is specific to the sphere. As soon as one deforms the surface away from the sphere and that the surface acquires a non-trivial shape, the traceless quadrupole tensor does not vanish anymore and provides a measure of how far the surface shape is from the spherical geometry. 
We can further introduce a shape scalar $\tau=\tr \,\tt^{2}=\tr\,t^{2}-(\tr\,t)^{2}$, which defines a rotation-invariant measure of the shape of the surface. It vanishes, $\tau=0$, if and only if the surface is a sphere.

\medskip

{\it Example 2: the Ellipsoid}

\noindent
Let us consider the simplest surface with a non-trivial quadrupole moment. We use the angular variables to parametrize the ellipsoid:
\be
\f{x^{2}}{a^{2}}+\f{y^{2}}{b^{2}}+\f{z^{2}}{c^{2}}=1\,,\qquad
\vx=\mat{c}{a\sin\theta\cos\phi \\ b\sin\theta\sin\phi \\ c\cos\theta}
\,,\quad
\theta\in[0,\pi]
\,,\quad
\phi\in[0,2\pi]
\,,
\ee
where $a,b,c$ are the three radii of the ellipsoid. The infinitesimal area element is slightly more complicated than for the sphere:
\be
\sqrt{\det g}=\sin\theta\,\sqrt{b^{2}c^{2}\sin^{2}\theta\cos^{2}\phi+c^{2}a^{2}\sin^{2}\theta\sin^{2}\phi+a^{2}b^{2}\cos^{2}\theta}.
\ee
When the three radii match, $a=b=c=R$, the square-root factor simplifies to $R^{2}$ leading to the correct area element for the sphere. This factor makes it harder to perform the surface integrations. For instance, both the ellipsoid area and the quadrupole moment  do not admit generic closed formula in terms of simple functions, and necessarily involves (incomplete) elliptic integrals. They nevertheless simplifies in the case of spheroids, when two radii match.

The ellipsoid area reads:
\be
\cA=\iint \sqrt{\det g}
=abc A(\alpha,\beta,\gamma)\,,
\quad\textrm{with}\quad
A(\alpha,\beta,\gamma)
=\int\dd \theta \dd \phi\,
\sqrt{\alpha \sin^{2}\theta\cos^{2}\phi+\beta\sin^{2}\theta\sin^{2}\phi+\gamma\cos^{2}\theta}\,,
\ee
with $\alpha=a^{-2}$, $\beta=b^{-2}$ and $\gamma=c^{-2}$. Its expression in terms of elliptic integrals is traditionally given by the Legendre formula, for $a\ge b \ge c$:
\be
\label{Areaellipsoid}
\cA=2\pi c^{2}+\f{2\pi ab}{\sin\vphi}\big{(}\sin^{2}\vphi\, E(\vphi,m)+\cos^{2}\vphi \,F(\vphi,m)\big{)}\,,
\quad
m=\f{1-\f{c^{2}}{b^{2}}}{1-\f{c^{2}}{a^{2}}}\,,
\quad
\cos\vphi=\f{c}{a}\,,
\ee
where the incomplete elliptic integrals are defined as:
$$
E(\vphi,m)=\int_{0}^{\vphi}\dd\phi\,\sqrt{1-m\sin^{2}\phi}
=\int_{0}^{\sin\vphi}\dd t\,\sqrt{\f{1-mt^{2}}{1-t^{2}}}\,,
$$
$$
F(\vphi,m)=\int_{0}^{\vphi}\f{\dd\phi}{\sqrt{1-m\sin^{2}\phi}}
=\int_{0}^{\sin\vphi}\f{\dd t}{({1-mt^{2}})({1-t^{2}})}\,.
$$
The curious reader will find a full list of alternative formulas in terms of elliptic integrals in \cite{Ellipsoid2006} and of hypergeometric functions in \cite{Rivin2007409,Ellipsoid2013}.
The quadrupole tensor gives similar integrals. All off-diagonal component still vanish but the matrix isn't proportional to the identity anymore.  The diagonal components are given by:
\be
\mat{c}{t_{11}\\t_{22}\\t_{33}}
=abc
\int\dd\theta\dd\phi\,
\sqrt{\alpha \sin^{2}\theta\cos^{2}\phi+\beta\sin^{2}\theta\sin^{2}\phi+\gamma\cos^{2}\theta}\,
\mat{c}{a^{2}\sin^{2}\theta\cos^{2}\phi \\b^{2}\sin^{2}\theta\sin^{2}\phi\\c^{2}\cos^{2}\theta}\,.
\ee
We could express these in terms of elliptic integrals, similarly to the ellipsoid area, but this leads only to clumsy formulae. We nevertheless point out that, although the trace of the quadrupole moment matrix $\tr\,t$ does not seem to reproduce any special geometrical observable, the weighted trace gives back the ellipsoid area:
\be
\f{t_{11}}{a^{2}}+\f{t_{22}}{b^{2}}+\f{t_{33}}{c^{2}}=\cA
\,.
\ee

\medskip

In the case of spheroids, or ellipsoids of revolution, when two radii match, these integrals can be explicitly performed. Let us chose $a=b$. Then in the case of a prolate ellipsoid, wtih $c>a=b$, we get:
\beq
\label{Aaac}
\cA_{a=b<c}&=&
2\pi a\int_{0}^{\pi}d\theta\,\sin\theta
\sqrt{(c^{2}-a^{2})\sin^{2}\theta+a^{2}}\nn\\
&=&
2\pi a^{2}
+\f{2\pi ac}{\sqrt{1-\f{a^{2}}{c^{2}}}}\arctan\sqrt{\f{c^{2}}{a^{2}}-1}
\,=\,
2\pi a^{2}
+\f{2\pi ac}{\sqrt{1-\f{a^{2}}{c^{2}}}}\arcsin\sqrt{1-\f{a^{2}}{c^{2}}}
\,.
\eeq
And the case of an oblate ellipsoid, with $c<a=b$, can be obtained as an analytical continuation of those formulae, which gives:
\beq
\cA_{a=b>c}&=&
2\pi a^{2}
+\f{2\pi ac}{\sqrt{\f{a^{2}}{c^{2}}-1}}\arctanh\sqrt{1-\f{c^{2}}{a^{2}}}
\,=\,
2\pi a^{2}
+\f{2\pi ac}{\sqrt{\f{a^{2}}{c^{2}}-1}}\arcsinh\sqrt{\f{a^{2}}{c^{2}}-1}
\,.
\eeq
\begin{figure}[h]
  \centerline{\includegraphics[height=4cm]{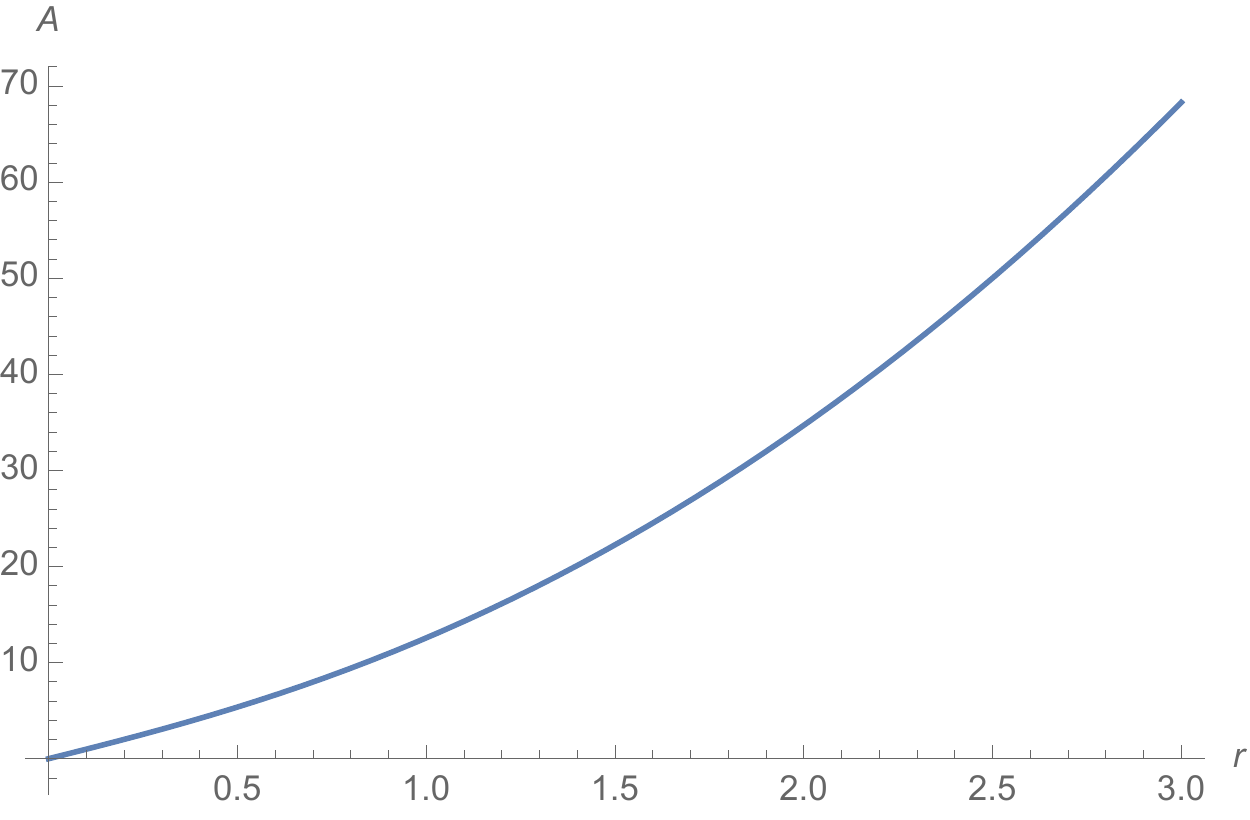}\hspace{2cm}\includegraphics[height=4cm]{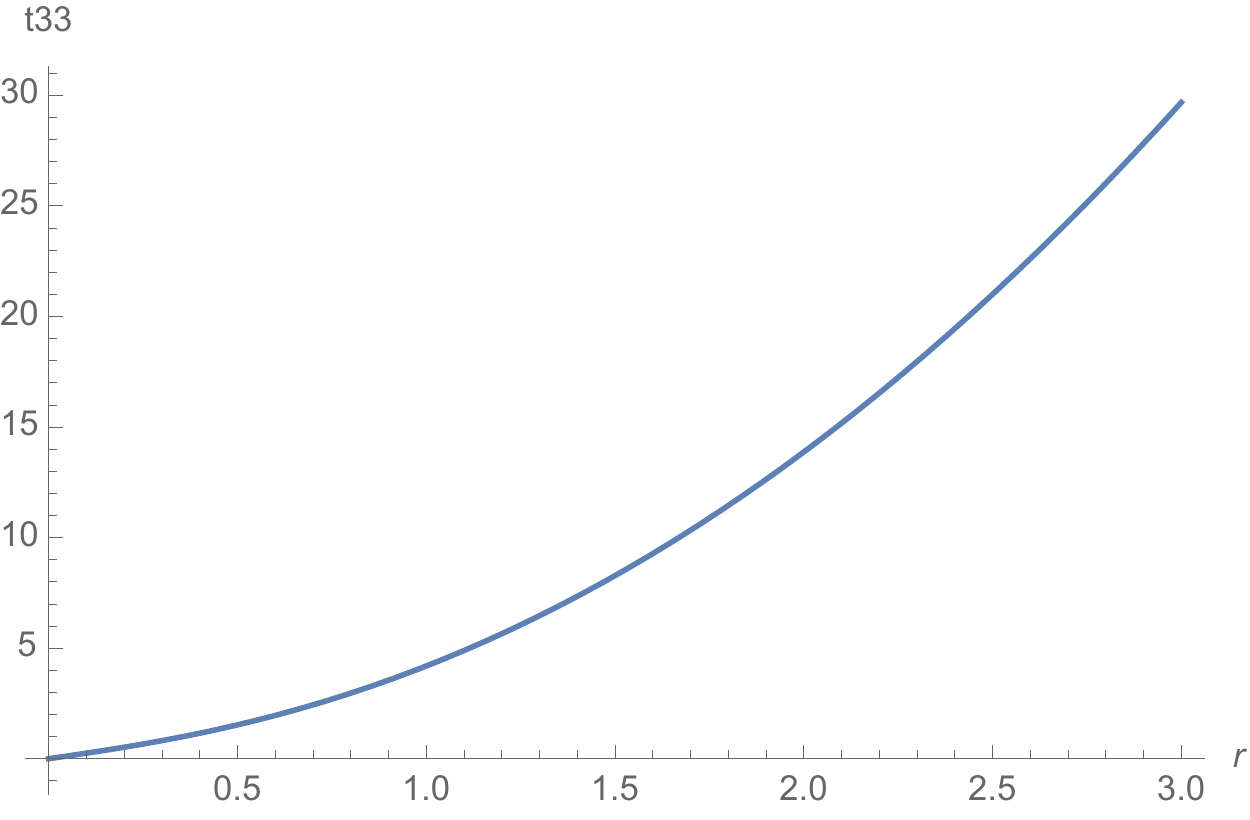}}
  \label{fig:plott33}
  \caption{Plots of the area $\cA$ (on the left) and the quadrupole moment $t_{33}$ (on the right) for an ellipsoid of revolution in terms of the ratio $r=a/c$ for $c=1$.}
\end{figure}

We similarly compute the diagonal components of the quadrupole matrix. Assuming that $a=b$, we get for the third component:
\be
t_{33}^{a=b,c}
=
2\pi a^{2}c^{2}\int_{0}^{\pi}d\theta\,\sin\theta\cos^{2}\theta
\sqrt{\f{c^{2}}{a^{2}}\sin^{2}\theta+\cos^{2}\theta}
\,=\,
\f\pi2 a^{2}c^{2}\,\left(\f{1-2\f{a^{2}}{c^{2}}}{1-\f{a^{2}}{c^{2}}}\right)
+\f\pi2 ac^{3}\,\f{\arctan\sqrt{\f{c^{2}}{a^{2}}-1}}{\left(1-\f{a^{2}}{c^{2}}\right)^{\f32}}
\,,
\ee
which behaves as, with the full plot given in fig.\ref{fig:plott33}:
\be
t_{33}^{a=b,c}\,\underset{c\ll a}\sim\pi a^{2}c^{2}\left(1+\f12\f{c^{2}}{a^{2}}\right)
\qquad
t_{33}^{a=b,c}\,\underset{c\arr a}\sim\f{4\pi}3a^{2}c^{2}+\f{8\pi}{15}a(c-a)c^{2}
\qquad
t_{33}^{a=b,c}\,\underset{c\gg a}\sim\f{\pi^{2}}4ac^{3}
\,.
\ee
From $t_{33}$, we can deduce the other components by the weighted trace identity:
\be
t_{11}^{a=b,c}=t_{22}^{a=b,c}=
\f{a^{2}}2\left(\cA-\f{t_{33}}{c^{2}}\right)
\,=\,
\f\pi4a^{4}\left(\f{3-2\f{a^{2}}{c^{2}}}{1-\f{a^{2}}{c^{2}}}\right)
+\f\pi4 a^{3}c\,
\left(\f{3-4\f{a^{2}}{c^{2}}}{1-\f{a^{2}}{c^{2}}}\right)
\f{\arctan\sqrt{\f{c^{2}}{a^{2}}-1}}{\left(1-\f{a^{2}}{c^{2}}\right)^{\f12}}\,.
\ee

Even if these integrals do not have a generic closed analytical expression for arbitrary values of the radii, there is a one-to-one correspondance between the three quadrupole moment eigenvalues $(t_{11},t_{22},t_{33})$ and the three principal radii of the ellipsoid $(a,b,c)$. Therefore, any closed surface (with trivial topology) can be approximated (up to second order in its moments) by an ellipsoid determined by its quadrupole moment: the rotation $o$ diagonalizing the quadrupole tensor $t$ determines the ellipsoid principal axis while the quadrupole eigenvalues determine the ellipsoid radii.

It might be interesting to compare the ellipsoid volume $V=4\pi abc/3$ with the volume $\cV$ of the region bounded by the original  surface and with the determinant $(\det t)^{\f14}$ of the quadrupole tensor. This might later provide a new approximation for the volume for discrete geometries.
One could also wonder if there is another family  of closed surface, different from the ellipsoid, which would lead computable surface integrals and a simpler expression of the quadrupole moment, but we haven't any.

\subsection{Surface Normal and Dual Quadrupole}
\label{app:dualellipsoid}

We now move on to the dual multipole moments, or equivalently to the multipole moments of the normal vector field defined as:
\be
\vN=\pp_{u}\vx \w \pp_{v}\vx,
\qquad
N= |\vN|,
\qquad
\hn=\f{\vN}{N}
\,,
\ee
The norm  gives the infinitesimal area element, $N= \sqrt{\det g}$, and the unit normal $\hn$ is invariant under reparametrization (it is a scalar), so we define the dual moments as the integral means of the powers of the unit normal.
The dual monopole moment co\"\i ncides with the standard monopole and gives the area, $\cM_{0}=\cI_{0}=\cA$.
The dual dipole is the closure vector (or defect) usually used in loop quantum gravity and twisted geometry:
\be
\cM_{1}^{a}
\,=\,
\int_{\cS} \dd u \dd v\,N \,n^{a}
\,=\,
\int_{\cS} \dd u \dd v\,N^{a}
=
\vcC\,.
\ee
When the surface is closed, as we have assumed here, the closure vector vanishes, $\cM_{1}=0$.
The dual quadrupole is the main object of interest of the present work:
\be
T^{ab}
\,=\,
\cM_{2}^{ab}
\,=\,
\int_{\cS}\dd u \dd v\,N \,n^{a}n^{b}
\,=\,
\int_{\cS} \dd u \dd v\,\f1N \,N^{a}N^{b}\,.
\ee
The trace of this dual quadrupole tensor is simply the surface area:
\be
\tr\,T=\int _{\cS} \dd u \dd v\,N=\cA\,,
\ee
so we can introduce the traceless part of the dual quadrupole moment, removing the information already contained by the lower order moments:
$$
\tT^{ab}=T^{ab}-\f13\tr T \,\id\,.
$$
Since the surface normals rotate as vectors under 3d rotations, one can rotate the surface in order to diagonalize this dual quadrupole tensor:
\be
T={}^{t}O \mat{ccc}{\mu_{1} & & \\ &\mu_{2}& \\ &&\mu_{3} }O\,,\quad
O\in\SO(3)
\ee
The change of basis matrix $O$ is a priori different from the matrix $o$ diagonalizing the original quadrupole moment, although they do match the simple case of an ellipsoid.
So for an arbitrary topologically-trivial closed surface, its dual quadrupole tensor defines a unique ellipsoid approximation, such that this ellipsoid yields the exact same dual quadrupole moment. The rotation $O$ defines the principal axes of the ellipsoid. 

The three eigenvalues $\mu_{a}$ are related to the three eigenvalues $\lambda_{a}$ of the original quadrupole moment and encode the  shape of the surface.  As we discuss below, we can extract the three ellipsoid radii from these three quadrupole matrix eigenvalues, and thus define the ellipsoid approximating the surface.
This ellipsoid approximation is a priori different from the one obtained from the usual quadrupole moment for the position. This new ellipsoid approximation defined by the dual quadrupole has exactly the same area as the original surface.

\bigskip

The components of the dual quadrupole can actually be understood as the response of the surface area with respect to its basic deformations. For instance, let us introduce the 3-parameter deformations of the surface by rescaling along the three axis:
\be
\vx\arr\mat{c}{ax^{1}\\bx^{2}\\ cx^{3}}\,,
\quad
\vN\arr\mat{c}{bc N_{1}\\ ca N_{2}\\ab N_{3}}\,,
\quad
\cA\arr \cA(a,b,c)
=\int \dd u \dd v\,
\sqrt{b^{2}c^{2} (N_{1})^{2}+ c^{2}a^{2} (N_{2})^{2}+a^{2}b^{2} (N_{3})^{2}}
\,\,.
\ee
Differentiating with respect to $a$ will produce a factor $N_{1}^{2}$ and send the square-root to the denominator, thus leading to the diagonal component $T_{11}$. Adjusting the $a,b,c$ factors, we get:
\be
T_{11}
\,=\,
-a^{3}\f{b^{2}c^{2}}{abc}\,\f{\pp }{\pp a}\left(\f{\cA}{abc}\right)
\,=\,
\cA-a\f{\pp \cA}{\pp a}
\,,
\ee
which we evaluate at $a=b=c=1$ to get the quadrupole of the original surface.
This formula for the diagonal components are actually compatible with the trace identity $\tr\, T=\cA$ giving the area as the trace of the dual quadrupole matrix:
\be
\cA=T_{11}+T_{22}+T_{33}=3\cA-\left(a\f{\pp \cA}{\pp a}+b\f{\pp \cA}{\pp b}+c\f{\pp \cA}{\pp c}\right)=3\cA-2\cA\,,
\ee
where the homogeneous dilatation operator $(a\pp_{a}+b\pp_{b}+c\pp_{c})$ simply  acts on the area, which is of length dimension 2.
We can similarly obtain the off-diagonal quadrupole components as area deformations by slightly skewing the dilatation in order to mix the position vector components. Let us consider the following new surface deformation:
\be
\vx\arr\mat{c}{x^{1}+\lambda x^{2}\\ x^{2} \\ x^{3}}
\,\quad
\vN\arr\mat{c}{N_{1}\\ N_{2}-\lambda N_{1} \\ N_{3}}
\,\quad
\cA\arr
\cA(\lambda)
=\int \dd u \dd v\,
\sqrt{(1+\lambda^{2})(N_{1})^{2}+ (N_{2})^{2}+ (N_{3})^{2}-2\lambda N_{1}N_{2}}\,,
\nn
\ee
which allows to extract the off-diagonal $T_{12}$ matrix element:
\be
T_{12}
\,=\,
\left[\lambda T_{11}-\f{\pp_{\lambda}\cA}{\pp\lambda}\right]_{\lambda=0}
\,=\,
-\left[\f{\pp_{\lambda}\cA}{\pp\lambda}\right]_{\lambda=0}
\,.
\ee
These differential expressions for the dual quadrupole moment show that it describes the shape of the surface but also reflects the basic deformation modes of the surface area.

\medskip

{\it Example 1: the  Sphere}

\noindent
For the unit sphere, the unit normals match exactly the position vectors:

\be
\vx=\mat{c}{\sin\theta\cos\phi \\ \sin\theta\sin\phi \\ \cos\theta}
\,,\quad
\vN=\sin\theta\,\mat{c}{\sin\theta\cos\phi \\ \sin\theta\sin\phi \\ \cos\theta}=\sin\theta\,\vx
\,,\quad
N=\sin\theta
\,,
\hn=\vx\,,
\ee
so that the original and dual moments match exactly: 
\be
\cM_{0}=\cI_{0}=4\pi=\cA
\,,\qquad
\cM_{1}=\cI_{1}=0
\,,\qquad
\cM_{2}=\cI_{2}=\f{4\pi}{3}\,\id=T\,.
\ee

\medskip

{\it Example 2: the Ellipsoid}

\noindent
The ellipsoid is the exact re-scaling of the unit sphere by dilatations along the 3 axis with ratio given by the three radii, $a$, $b$ and $c$. We can then apply the differential formulae above to express the dual quadrupole tensor as derivatives of the area $\cA$ as a function of $a,b,c$. The off-diagonal components vanish and the diagonal components are given by:
$$
T_{11}=\cA-a\f{\pp \cA}{\pp a}
\,,\qquad
T_{22}=\cA-b\f{\pp \cA}{\pp b}
\,,\qquad
T_{33}=\cA-c\f{\pp \cA}{\pp c}\,,
$$
where $\cA(a,b,c)$ is given by eqn.\eqref{Areaellipsoid} in the previous section in terms of  elliptic integrals. These three eigenvalues of the dual quadrupole tensor will thus also given in terms of incomplete elliptic integrals and will not admit an expression in terms of simpler functions.

\begin{figure}
  \centerline{\includegraphics[height=3cm]{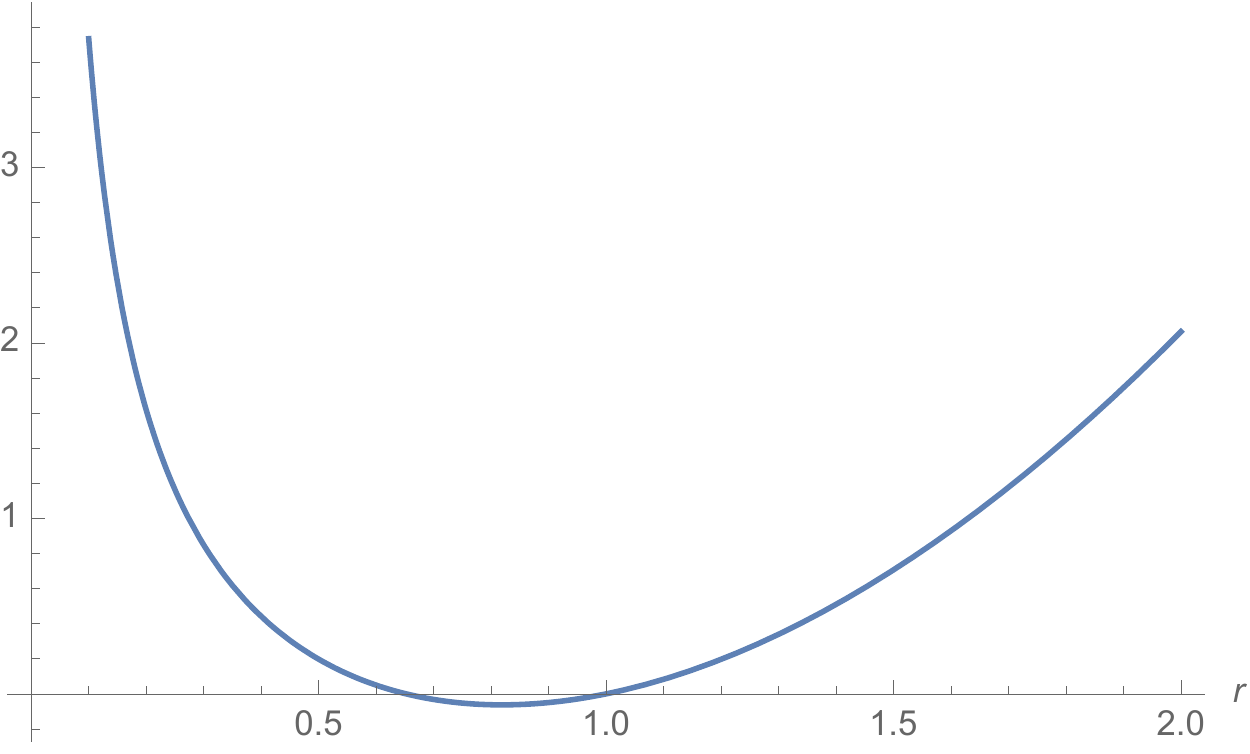}\hspace{8mm}\includegraphics[height=3cm]{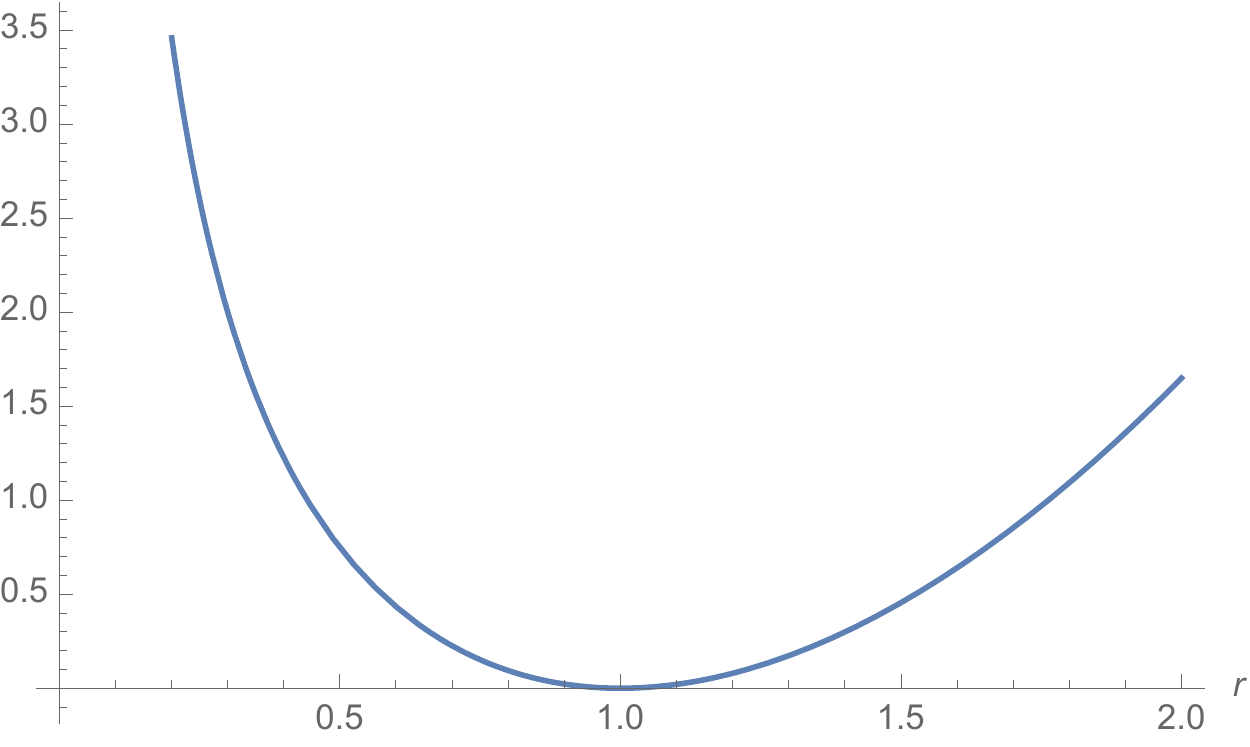}\hspace{8mm}\includegraphics[height=3cm]{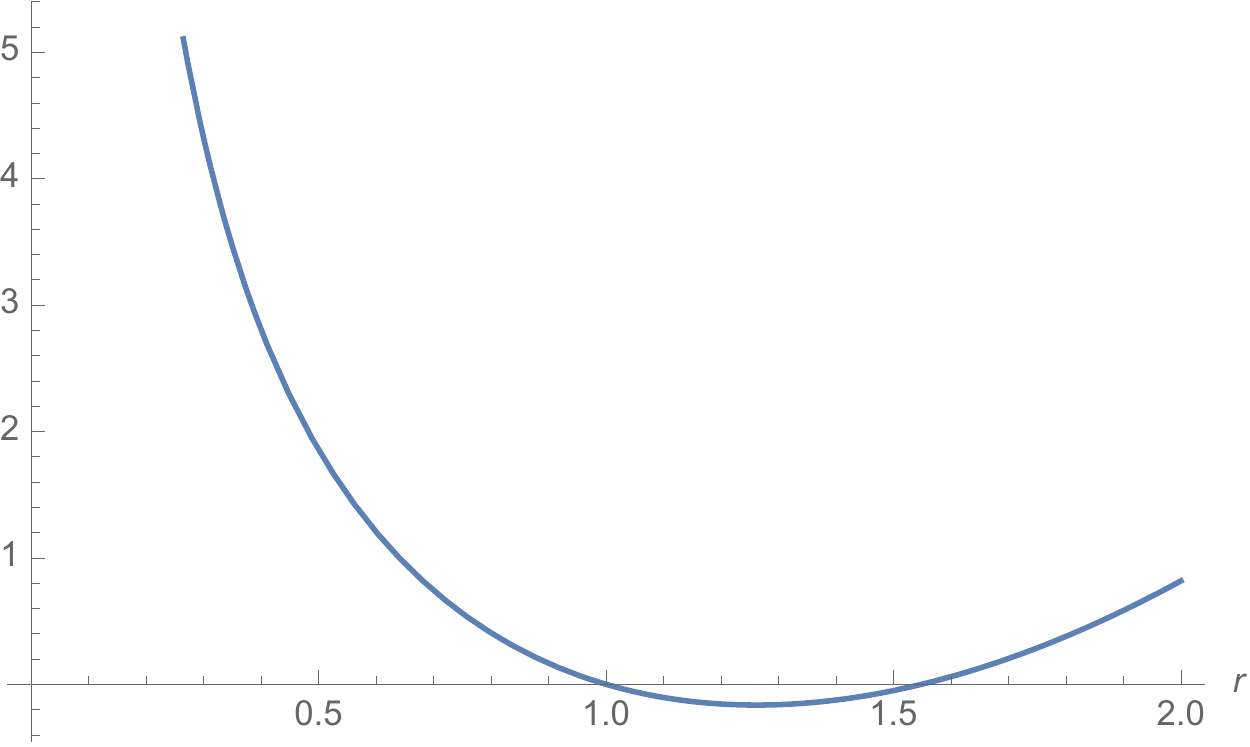}}
  \label{fig:plotr}
  \caption{Plots of the condition \eqref{r-eqn} for the ratio $r=a/c$ for various values of the quadrupole ratio $\tau=T_{33}/T_{11}$: from the left to the right, $\tau=1/2$, $\tau=1$ and $\tau=2$. The equation on $r$ always has two positive roots, except for the isotropic case $\tau=1$ which leads back to the sphere $r=1$ as expected. }
\end{figure}

As before, the integrals simplify for an ellipsoid of revolution, when two radii match, say $a=b$. Starting from the area formula \eqref{Aaac}, we differentiate with respect to the third radius $c$ to get the component $T_{33}$ then use the trace identity to get $T_{11}$. This gives (assuming a priori that $c>a$, although all the expressions are finite and obviously analytically continued beyond this bound):
\be
T_{11}=\pi\f{a}{1-\f{a^{2}}{c^{2}}}\,\left[
a+\f{c\left(1-\f{2a^{2}}{c^{2}}\right)}{\sqrt{1-\f{a^{2}}{c^{2}}}}\arcsin{\sqrt{1-\f{a^{2}}{c^{2}}}}
\right]\,,
\ee
\be
T_{33}=
\f{a^{2}}{c^{2}}2\pi\f{a}{1-\f{a^{2}}{c^{2}}}\,\left[
-a+\f{c}{\sqrt{1-\f{a^{2}}{c^{2}}}}\arcsin{\sqrt{1-\f{a^{2}}{c^{2}}}}
\right]\,,
\ee
\be
\cA=2T_{11}+T_{33}=
2\pi a\,\left[
a+\f{c}{\sqrt{1-\f{a^{2}}{c^{2}}}}\arcsin{\sqrt{1-\f{a^{2}}{c^{2}}}}\right]
\nn
\ee
We can combine $T_{11}$ and $T_{33}$ in two different ways to get interesting identities:
\be
2T_{11}+\left(2-\f1{r^{2}}\right)T_{33}
\,=\,
4\pi a^{2}
\,=\,
\f{\sqrt{1-r^{2}}}{r\arcsin\sqrt{1-r^{2}}}
\left[2r^{2}T_{11}+T_{33}\right]
\quad\textrm{with}\quad
r\equiv\f{a}c\,.
\ee
This allows to solve the inverse problem for the ellipsoid of revolution. Let us start with the dual quadrupole moments $T_{11}$ and $T_{33}$. We use the identity to solve for the ratio $r=a/c$ in terms of the quadrupole values:
\be
\label{r-eqn}
\left[2+\left(2-\f1{r^{2}}\right) \tau\right]
-
\f{\sqrt{1-r^{2}}}{r\arcsin\sqrt{1-r^{2}}}
\left[2r^{2}+\tau\right]
\,=\,0
\quad\textrm{with}\quad
\tau\equiv\f{T_{33}}{T_{11}}\,.
\ee
Assuming a value $\tau>0$, this equation generically has two real positive roots for $r\in\R^{+}$. First, whatever the value of $\tau$, the point $r=1$ is always a root, but this is generically a fiducial solution. For a unit ratio $\tau=1$, then $r=1$ is indeed the only and the correct solution, and we recover the sphere. As shown on fig.\ref{fig:plotr}, when $\tau<1$, we get another root $0<r<1$. And when $\tau>1$, we similarly get another root $r>1$. Once this equation for the radii ratio $r$ is solved, at least numerically, we can get $a$ by the identity given above and finally deduce $c$.
In general, it is also possible to invert numerically for the three radii $a,b,c$ from the three quadrupole moments $T_{11},T_{22},T_{33}$.

\section{Shape of a Tetrahedron}

\subsection{Reconstruction of the normal vectors of the tetrahedron in the (area,angle) parametrization}
\label{ap:tetrahedron_angle_area_param}

\begin{figure}[h!]
	\begin{center}
		\begin{tikzpicture}[scale=1.2]
		
		\coordinate (O) at (10,2);
		\coordinate (x) at (13,2);
		\coordinate (z) at (10,5.5);
		\coordinate (y) at (10.8,2.6);
		
		\coordinate (N1) at (10.8,3.6);
		\coordinate (N2) at (9.2,3.8);
		\coordinate(N2') at (10,5.2);
		
		\coordinate (N3) at (9.3,0.6);
		\coordinate (N4) at (11.2,0.8);
		
		\coordinate (varphiR) at (13,2.7);
		\coordinate (varphiL) at (8,1.5);
		
		\fill[color=orange!20] (varphiL)--(varphiR) --+(0,-1.5)--+(-5,-3)--cycle;
		\fill[color=red!20] (x)--+(0,4)--+(-5,4)--+(-5,0)--cycle;
		\fill[pattern=north west lines, pattern color=orange!70] (x)--(O)--(varphiR)--cycle;
		
		
		\draw[dashed, thick] (varphiR)--(varphiL);
		\draw[thick] (11.5,2) arc (-40:-15:1); \draw (11.6,2) node[above right]{$\varphi$};
		\draw[->] (10,2.5) arc (90:63:0.5); \draw (9.98,2.5) node[scale=0.75,above right]{$\theta_{12}$};
		\draw[->] (9.65,2.8) arc(110:66:1); \draw (9.8,2.85) node[scale=0.75,above]{$\alpha_{12}$};

		
		\draw[->] (O)-- node[pos=1,right]{$\vec{x}$}(x);
		\draw[->] (O)-- node[pos=1,above]{$\vec{y}$}(y);
		\draw[->] (O)-- node[pos=1,above]{$\vec{z}$}(z);
		
		\draw[red,->] (O)-- node[scale=0.7,pos=0.7,above left]{$\vec{N}_1$}(N1);
		\draw[color=red,->] (O)-- node[scale=0.7,pos=0.7,left]{$\vec{N}_2$}(N2);
		\draw[dashed,color=red,->] (N1)--  (N2');
		\draw[dashed,color=red,->] (N2)--  (N2');
		
		\draw[blue,->] (O)-- node[pos=0.55,left,scale=0.7]{$N_{12} \vec{z}$}(N2');
		
		\draw[orange,->] (O)-- node[pos=0.7,left]{$\vec{N}_3$}(N3);
		\draw[orange,->] (O)-- node[pos=0.7,right]{$\vec{N}_4$}(N4);
			
		
		\end{tikzpicture}
	\end{center}
	\caption{Dual representation of the tetrahedron: $\vec{N}_1$ and $\vec{N}_2$ lie on the $(xOz)$ plan, such that their sums is along $(Oz)$. $\varphi$ is the dihedral angle between the planes spanned by $(\vec{N}_1,\vec{N}_2)$ and $(\vec{N}_3,\vec{N}_4)$ }
	\label{ap:normals}
\end{figure}
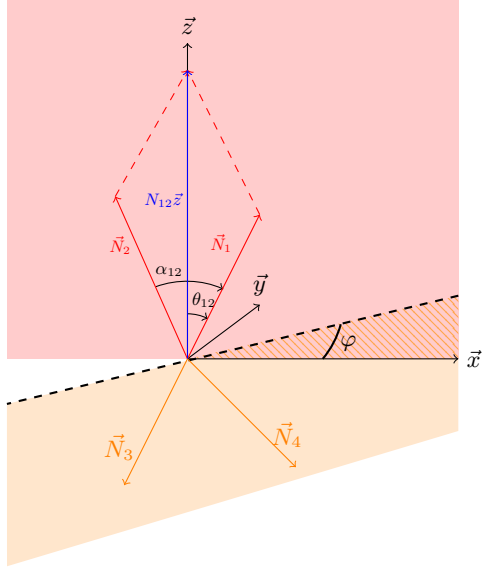

In order to reconstruct the normal vectors to a tetrahedron from the rotation-invariant data of its triangle areas $N_{i}$ and the canonical pair $(N_{12},\vphi)$ of the Kapovich-Millson symplectic structure, we need to fix the action of 3d rotations on tetrahedra. We choose the two vectors $\vN_{1}$ and $\vN_{2}$ to be in the  $(xOz)$ plane and their sum $\vN_{12}=\vN_{1}+\vN_{2}$ to go along the axis $(Oz)$, as illustrated on fig.\ref{ap:normals}.

Mathematically, we write:
\begin{equation}
\vec{N}_{12} = N_{12} \hat{z}= (\vec{N_1}+\vec{N}_2)=- (\vec{N}_3+\vec{N}_4)
\end{equation}
From the norm of $N_{12}$, we deduce the angle $\alpha_{12}$ (resp. $\alpha_{34}$) between $\vec{N}_1$ and $\vec{N}_2$ (resp. $\vec{N}_3$ and $\vec{N}_4$):
\begin{equation}
	\cos \alpha_{12} = \frac{N_{12}^2-N_1^2-N_2^2}{2 N_1 N_2}
	\,,\qquad
	\cos \alpha_{34} = \frac{N_{12}^2-N_3^2-N_4^2}{2 N_3 N_4}.
\end{equation}
Then we write  the vectors $\vN_{i}$ in terms of two angles $\theta_{12}$ and $\theta_{34}$:
\begin{align}
\vecc{N}_{1} &= N_1 \begin{pmatrix}
\sin\theta_{12}\\
0 \\
\cos\theta_{12}
\end{pmatrix}
,\; \qquad \quad \; \;
\vecc{N}_{2} = N_2 \begin{pmatrix}
\sin(\theta_{12}-\alpha_{12}) \\
0 \\
\cos(\theta_{12}-\alpha_{12})
\end{pmatrix} 
,\;
\\
\vecc{N}_3 &= - N_3 \begin{pmatrix}
\sin\theta_{34} \cos\varphi \\ 
\sin\theta_{34} \sin\varphi \\
\cos\theta_{34}
\end{pmatrix}
,\;
\vecc{N}_4 = - N_4 \begin{pmatrix}
\sin(\theta_{34}-\alpha_{34}) \cos(\varphi) \\ 
\sin(\theta_{34}-\alpha_{34}) \sin(\varphi) \\
\cos(\theta_{34}-\alpha_{34}) 
\end{pmatrix} \nonumber
\end{align}
Imposing that  $\vN_{12}\in (Oz)$ gives a constraint allowing to determine the angles $\theta_{12}$ and $\theta_{34}$:
\be
\left|
\begin{array}{ll}
	N_{1}\sin(\theta_{12}) + N_{2}\sin(\theta_{12}-\alpha_{12}) &= 0 \\
	N_{3}\sin(\theta_{34}) + N_{4} \sin(\theta_{34}-\alpha_{34}) &= 0
\end{array}
\right.
\qquad\Longrightarrow\quad
\tan\theta_{12} = \frac{N_2 s_{\alpha_{12}}}{N_1 + N_2 c_{\alpha_{12}}}
\,,\qquad
\tan\theta_{34} = \frac{N_4 s_{\alpha_{34}}}{N_3 + N_4 c_{\alpha_{34}}}
\,.
\ee
Combining everything together reconstruct the normals function of $(N_1,N_2,N_3,N_4,N_{12},\varphi)$.
In the iso-area case, the expression of the normals simplifies a lot and gives:
\begin{equation}
\vN_{1} = N \begin{pmatrix}
\sin\frac{\alpha}{2} \\
0 \\
\cos\frac{\alpha}{2} 
\end{pmatrix}
,\;
\vN_{2} = N \begin{pmatrix}
-\sin\frac{\alpha}{2} \\
0 \\
\cos\frac{\alpha}{2}
\end{pmatrix} 
,\;
\vN_3 = N \begin{pmatrix}
-\sin\frac{\alpha}{2} \cos\varphi \\ 
-\sin\frac{\alpha}{2} \sin\varphi \\
-\cos\frac{\alpha}{2}
\end{pmatrix}
,\;
\vN_4 =  N \begin{pmatrix}
\sin\frac{\alpha}{2} \cos\varphi \\ 
\sin\frac{\alpha}{2} \sin\varphi \\
-\cos\frac{\alpha}{2}
\end{pmatrix}
\quad\textrm{with}\,\,
\cos\f\alpha2=\f{N_{12}}{2N}\,.
\nn
\end{equation}

\subsection{Plots of the Jacobian for the change of variables $(N_{12},\vphi)\,\rightarrow\,(\tr\,\tT^{2}\,,\,\det\,T)$}
\label{ap:jacobian}

\begin{figure}[h]
	\includegraphics[height=40mm]{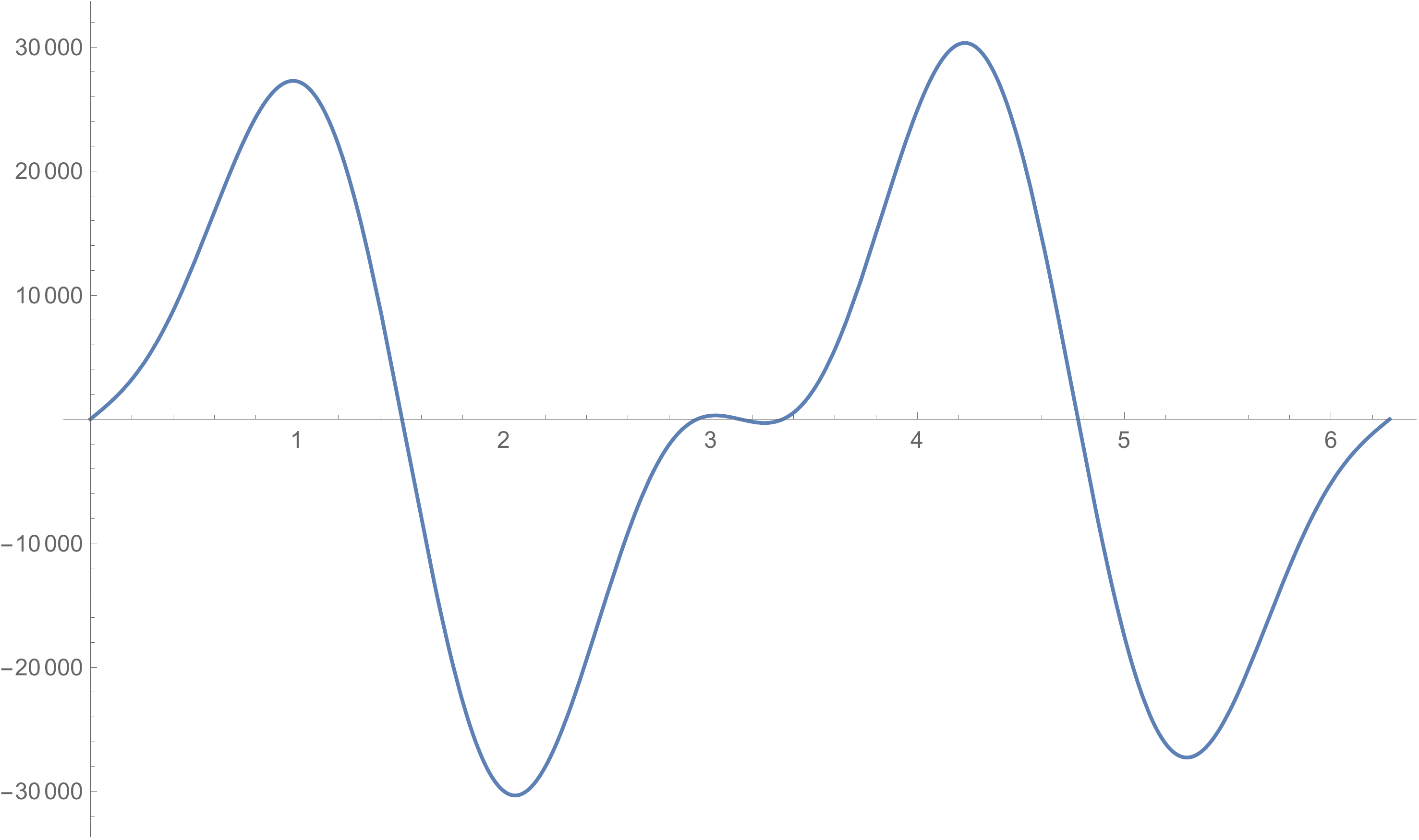}
\hspace{20mm}
	\includegraphics[height=40mm]{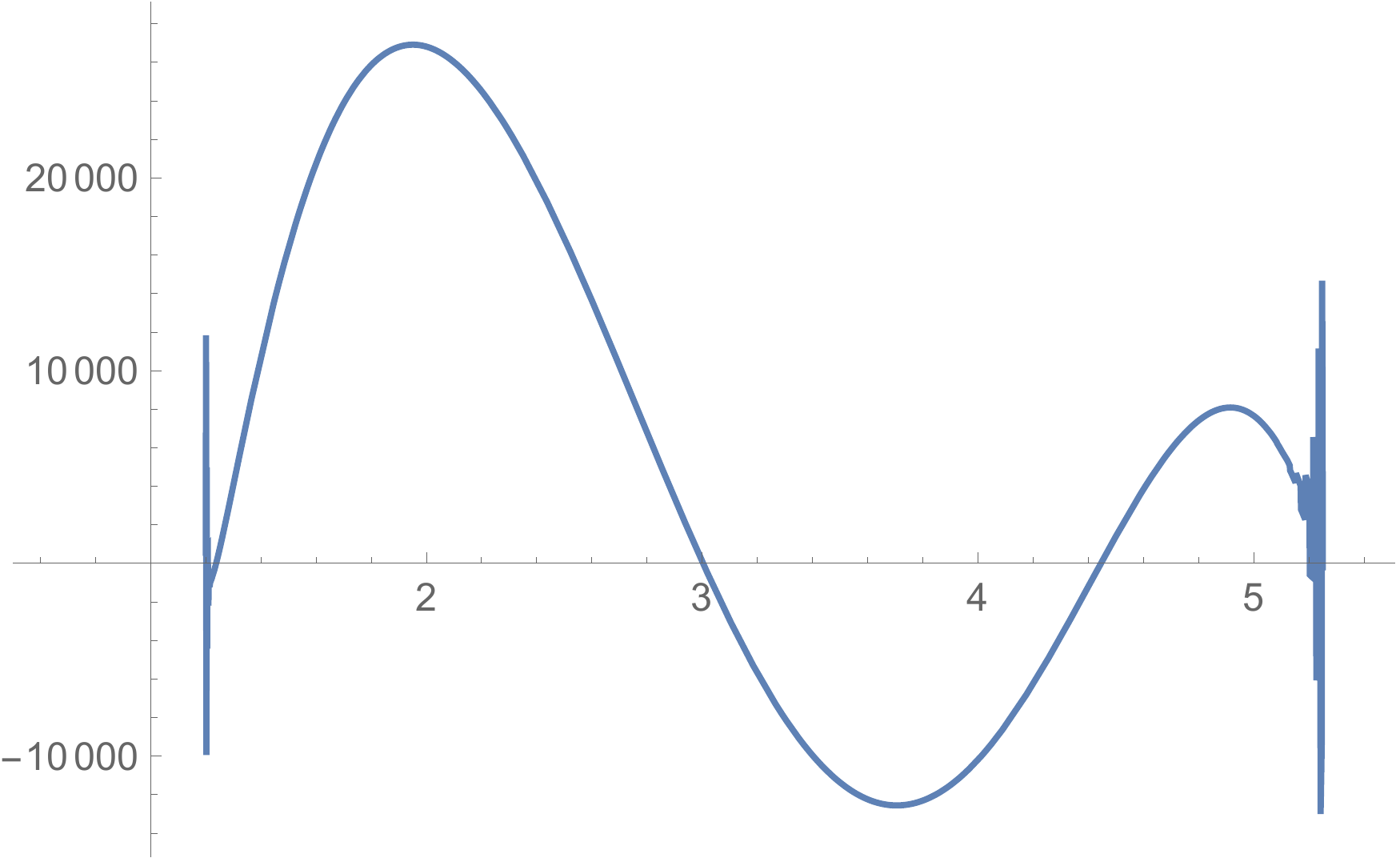}
	
	\caption{Plots of  the Jacobian of the change of variables $(N_{12},\varphi) \rightarrow (\det\,T,\tr\,\tT^2)$ for the area data  $(N_1,N_{2},N_{3},N_{4})=(7,9.4,5.4,5.1)$. On the right hand side, the Jacobian is drawn in terms of $\vphi$ for a fixed value $N_{12}=4$, while the plot on the left hand side gives the evolution of the Jacobian in terms of $N_{12}$ while $\varphi = \frac{\pi}{3} $ is held fixed.}
	\label{fig:jacobian}
\end{figure}

Here we plot the Jacobian of the change of variables  $(N_{12},\varphi)\rightarrow(\det\,T , \tr\,\tT^2)$  at fixed triangle areas. The goal is to show that the volume and shape observables, $\det\,T , \tr\,\tT^2$, uniquely determine the tetrahedron, similarly to the canonical pair $(N_{12},\varphi)$.
As the plots on fig.\ref{fig:jacobian} illustrate, this Jacobian determinant only admit isolated zeroes.


\bibliographystyle{bib-style}
\bibliography{surface}

\end{document}